\documentclass[twocolumn]{aastex631}
\usepackage{longtable}
\usepackage{amsmath}
\usepackage{mathrsfs}
\usepackage{lipsum}
\usepackage{placeins}
\usepackage{balance}

\newcommand{\htwo}{\ion{H}{2}}

\newcommand{\source}{ASKAP~J0055--2558}

\begin{document}

\title{ASKAP~J005512.2$–$255834: A Luminous, Long-Lived Radio Transient at $z = 0.1$ $-$ an Orphan Afterglow or an off-nuclear TDE from an IMBH?}

\correspondingauthor{Ashna Gulati}
\email{agul8829@uni.sydney.edu.au}

\author[0000-0002-0786-7307]{Ashna Gulati}
\affiliation{Sydney Institute for Astronomy, School of Physics, The University of Sydney, NSW 2006, Australia}
\affiliation{ARC Centre of Excellence for Gravitational Wave Discovery (OzGrav), Australia}
\affiliation{Australia Telescope National Facility, CSIRO, Space and Astronomy, PO Box 76, Epping, NSW 1710, Australia}

\author[0000-0002-2686-438X]{Tara Murphy}
\affiliation{Sydney Institute for Astronomy, School of Physics, The University of Sydney, NSW 2006, Australia}
\affiliation{ARC Centre of Excellence for Gravitational Wave Discovery (OzGrav), Australia}

\author[0000-0001-6295-2881]{David L. Kaplan}
\affiliation{Department of Physics, University of Wisconsin-Milwaukee, P.O. Box 413, Milwaukee, WI 53201, USA}
\affiliation{ARC Centre of Excellence for Gravitational Wave Discovery (OzGrav), Australia}

\author[0000-0003-0699-7019]{Dougal Dobie}
\affiliation{Sydney Institute for Astronomy, School of Physics, The University of Sydney, NSW 2006, Australia}
\affiliation{ARC Centre of Excellence for Gravitational Wave Discovery (OzGrav), Australia}

\author[0000-0002-4557-6682]{Charlotte Ward}
\affil{Department of Astronomy \& Astrophysics, 525 Davey Lab, 251 Pollock Road, The Pennsylvania State University, University Park, PA 16802, USA} 


\author[0000-0001-6544-8007]{Gemma Anderson}
\affiliation{Australia Telescope National Facility, CSIRO, Space and Astronomy, PO Box 76, Epping, NSW 1710, Australia}
\affiliation{Sydney Institute for Astronomy, School of Physics, The University of Sydney, NSW 2006, Australia}

\author[0000-0002-4079-4648]{Manisha Caleb}
\affiliation{Sydney Institute for Astronomy, School of Physics, The University of Sydney, NSW 2006, Australia}
\affiliation{ARC Centre of Excellence for Gravitational Wave Discovery (OzGrav), Australia}

\author[0000-0002-0844-6563]{Poonam Chandra}
\affiliation{National Radio Astronomy Observatory, 520 Edgemont Rd, Charlottesville, VA 22903, USA}

\author[0000-0001-5703-2108]{Jeff Cooke}
\affiliation{Centre for Astrophysics and Supercomputing, Swinburne University of Technology, Hawthorn, VIC 3122, Australia}
\affiliation{ARC Centre of Excellence for Gravitational Wave Discovery (OzGrav), Australia}

\author[0000-0001-8704-1822]{Barnali Das}
\affiliation{Australia Telescope National Facility, CSIRO, Space and Astronomy, P.O. Box 1130, Bentley WA 6102, Australia}

\author[0000-0001-9434-3837]{Adam Deller}
\affiliation{Centre for Astrophysics and Supercomputing, Swinburne University of Technology, Hawthorn, VIC 3122, Australia}
\affiliation{ARC Centre of Excellence for Gravitational Wave Discovery (OzGrav), Australia}

\author[0000-0003-3441-8299]{Adelle Goodwin}
\affiliation{International Centre for Radio Astronomy Research, Curtin University, Bentley, WA, Australia}

\author[0000-0002-0152-1129]{Kelly Gourdji}
\affiliation{Australia Telescope National Facility, CSIRO, Space and Astronomy, PO Box 76, Epping, NSW 1710, Australia}
\affiliation{ARC Centre of Excellence for Gravitational Wave Discovery (OzGrav), Australia}

\author[0000-0001-5876-9259]{Giancarlo Ghirlanda}
\affiliation{INAF - Osservatorio Astronomico di Brera, via E. Bianchi 46, I-23807 Merate (LC), Italy}
\affiliation{INFN - Sezione di Milano–Bicocca, Piazza della Scienza 3, 20126 Milano (MI), Italy}

\author[0000-0002-9994-1593]{Emil Lenc}
\affiliation{Australia Telescope National Facility, CSIRO, Space and Astronomy, PO Box 76, Epping, NSW 1710, Australia}

\author[0000-0001-8211-8608]{Anais M{\"o}ller}
\affiliation{Centre for Astrophysics and Supercomputing, Swinburne University of Technology, Hawthorn, VIC 3122, Australia}
\affiliation{ARC Centre of Excellence for Gravitational Wave Discovery (OzGrav), Australia}

\author[0000-0002-9415-3766]{James K.\ Leung}
\affiliation{David A. Dunlap Department of Astronomy and Astrophysics, University of Toronto, 50 St. George Street, Toronto, ON M5S 3H4, Canada}
\affiliation{Dunlap Institute for Astronomy and Astrophysics, University of Toronto, 50 St. George Street, Toronto, ON M5S 3H4, Canada}
\affiliation{Racah Institute of Physics, The Hebrew University of Jerusalem, Jerusalem 91904, Israel}

\author[0000-0002-4941-5333]{Stella Koch Ocker}
\affiliation{Cahill Centre for Astronomy and Astrophysics, California Institute of Technology, Pasadena, CA 91125 USA}
\affiliation{Observatories of the Carnegie Institution for Science, Pasadena, CA 91101 USA}

\author[0000-0003-1575-5249]{Joshua Pritchard}
\affiliation{Australia Telescope National Facility, CSIRO, Space and Astronomy, PO Box 76, Epping, NSW 1710, Australia}

\author[0000-0001-5231-2645]{Claudio Ricci}
\affiliation{Department of Astronomy, University of Geneva, ch. d’Ecogia 16, 1290, Versoix, Switzerland}
\affiliation{Instituto de Estudios Astrofísicos, Facultad de Ingeniería y Ciencias, Universidad Diego Portales, Av. Ejército Libertador 441, Santiago, Chile}

\author[0000-0002-1136-2555]{Elaine M. Sadler}
\affiliation{Sydney Institute for Astronomy, School of Physics, The University of Sydney, NSW 2006, Australia}
\affiliation{ARC Centre of Excellence for Gravitational Wave Discovery (OzGrav), Australia}
\affiliation{Australia Telescope National Facility, CSIRO, Space and Astronomy, PO Box 76, Epping, NSW 1710, Australia}

\author[0000-0003-4924-7322]{Om Sharan Salafia}
\affiliation{INAF - Osservatorio Astronomico di Brera, via E. Bianchi 46, I-23807 Merate (LC), Italy}
\affiliation{INFN - Sezione di Milano–Bicocca, Piazza della Scienza 3, 20126 Milano (MI), Italy}

\author[0009-0003-7685-183X]{Kavya Shaji}
\affiliation{Sydney Institute for Astronomy, School of Physics, The University of Sydney, NSW 2006, Australia}
\affiliation{ARC Centre of Excellence for Gravitational Wave Discovery (OzGrav), Australia}
\affiliation{Australia Telescope National Facility, CSIRO, Space and Astronomy, PO Box 76, Epping, NSW 1710, Australia}

\author[0000-0002-4622-796X]{Roberto Soria}
\affiliation{INAF - Osservatorio Astrofisico di Torino, Strada Osservatorio 20, I-10025 Pino Torinese, Italy}
\affiliation{Sydney Institute for Astronomy, School of Physics, The University of Sydney, NSW 2006, Australia}

\author{Mark Suhr}
\affiliation{Centre for Astrophysics and Supercomputing, Swinburne University of Technology, Hawthorn, VIC 3122, Australia}

\author{Artem Tuntsov}
\affiliation{Manly Astrophysics, 15/41-42 East Esplanade, Manly 2095, Australia}

\author[0000-0002-2066-9823]{Ziteng Wang} 
\affil{International Centre for Radio Astronomy Research, Curtin University, Bentley, Western Australia, Australia}

\begin{abstract}
We report the discovery of a slowly evolving, extragalactic radio transient, ASKAP J005512.2--255834 (hereafter \source), identified using the Australian SKA Pathfinder in a search for orphan afterglows associated with archival gravitational wave events. Although discovered in this context, there is no evidence that the transient is associated with any known gravitational wave event. Nonetheless, this source exhibits a 20-fold increase in flux density over $<250$ days, and it remains in a declining, detectable state more than 1000 days after the initial detection. Follow-up observations from 0.3 to 9~GHz reveal an evolving spectrum consistent with synchrotron emission. \source\ is spatially coincident with a low-mass, star-forming galaxy at redshift $z = 0.116$ ($d_{\rm L}$= 543 Mpc), placing its peak radio luminosity at $\nu L_\nu \sim 10^{39}\,\rm erg\,s^{-1}$. Analysis of its radio light curve, inferred blastwave velocity, energetics, host galaxy properties and the absence of counterparts at other wavelengths suggest that \source\ is most consistent with either the late-time phase of an orphan long gamma-ray burst afterglow or a tidal disruption event involving an intermediate-mass black hole spatially offset from the galaxy nucleus. The radio discovery of either of these phenomena is extremely rare, with only a few or no confirmed examples to date.

\end{abstract}

\keywords{High energy astrophysics(739) --- Radio transient sources(2008) --- Transient sources(1851) --- Gamma-ray bursts(629) --- Tidal disruption(1696)}

\section{Introduction} 
\label{sec:intro}

The study of explosive transients at radio wavelengths has expanded significantly in recent years, driven by large surveys using instruments such as the Australian SKA Pathfinder  \citep[ASKAP;][]{2007PASA...24..174J,hotan21}
and the Karl G. Jansky Very Large Array \citep[VLA;][]{2011ApJ...739L...1P}. Until recently, the discovery of explosive transients such as gamma-ray bursts (GRBs), supernovae (SNe), and tidal disruption events (TDEs) has largely relied on high-energy telescopes such as Swift and Fermi Gamma-ray Burst Monitor (GBM), and optical survey facilities such as the Zwicky Transient Facility (ZTF) \citep{2004ApJ...611.1005G,2009ApJ...702..791M,2019PASP..131a8002B}. While these will remain essential for studying prompt emission and identifying the transient type, they can miss events that are either dust obscured \citep{2012ApJ...756..111M} or beamed away from the observer \citep{2003ApJ...591.1097R,2001ApJ...562L..55F}. Radio observations are well-suited to detecting shock-driven emission produced as relativistic or sub-relativistic outflows \citep{1982ApJ...259..302C,1997ApJ...476..232M} interact with the surrounding medium, even in cases where the event is viewed off-axis \citep{2015A&A...578A..71G}. This makes radio wavelengths a critical tool for discovering transients and measuring rates for asymmetric transients with reduced selection bias.

Earlier radio surveys were constrained by limited sensitivity and cadence, making them ineffective at detecting extragalactic transients \citep[e.g.][]{2002ApJ...576..923L,2006ApJ...639..331G,2010ApJ...719...45C,2011MNRAS.415....2B,2011MNRAS.418.2813B,2016MNRAS.458.3506R, 2017MNRAS.466.1944M}. Instead, some transient searches focused on deep and targeted radio follow-up of nearby, face on spiral galaxies, with some yeilding detections of long-lived synchrotron transients that remain unclassified \citep[e.g.][]{2010MNRAS.404L.109M,2011MNRAS.415L..59J,2017ApJ...841..117P,2019MNRAS.486.3388D,2020ApJ...901L..17T,2019MNRAS.489.1181A}. Recent wide-field surveys, such as the Variables and Slow Transients Survey \citep[VAST;][]{2021PASA...38...54M}, the Rapid ASKAP Continuum Survey \citep[RACS;][]{2020PASA...37...48M} and the Very Large Array Sky Survey \citep[VLASS;][]{2020PASP..132c5001L} have significantly enhanced our ability to detect a range of synchrotron transient classes at low frequencies such as orphan afterglows  \citep{2023MNRAS.523.4029L}, off-axis TDEs \citep{2024ApJ...973..104D}, changing-look AGN \citep[CL-AGN;][]{2025AAS...24614216B},  and rebrightening supernovae \citep{rose_late-time_2024}.

In this paper, we present ASKAP~J005512.2--255834 (hereafter \source), identified as a luminous, slowly evolving radio transient, in a search for orphan afterglows from archival gravitational wave events \citep[see][for details]{2025MNRAS.538.2676G}. In Section~\ref{s_discovery}, we describe the discovery of the transient. Section~\ref{sec:data} details the multiwavelength observation campaign. In Section~\ref{s_transient_identification}, we describe its basic properties. Section~\ref{sec:host_galaxy} focuses on the analysis of the candidate host galaxy. In Section~\ref{sec:classification}, we summarise the key observational properties of the transient and attempt its classification by comparison with known classes of astrophysical transients. In Section~\ref{sec:physical_plausibility}, we evaluate the plausibility of interpreting this event as either an orphan afterglow or an off-nuclear TDE. Finally, we draw our conclusions in Section~\ref{sec:conc}. Throughout this work, we assume a flat $\Lambda$CDM cosmology with $H_0 = 69.6~\mathrm{km}~\mathrm{s}^{-1}~\mathrm{Mpc}^{-1}$, $\Omega_{\mathrm{M}} = 0.286$, and $\Omega_{\Lambda} = 0.714$ \citep{2013ApJS..208...19H}. 

\section{Discovery}
\label{s_discovery}

We identified \source\ in archival ASKAP data during a search for off-axis afterglows in the GW190814 localisation region. The search was conducted to a survey sensitivity $F_{\nu,\mathrm{lim}}$, range of 0.15 -- 1.5\,mJy (most constraining limit from a dedicated ASKAP project AS111 and least constraining from RACS), over an effective area of 22\,deg$^2$ corresponding to the 90\% localisation region of GW190814 \citep[more comprehensive search details given in][]{2025MNRAS.538.2676G}. The source is absent in earlier epochs and first appears on 2022 February 4 at 3 GHz, at least $\sim 20\times$ brighter than the last non-detection limit $250$ days prior at 0.9 GHz (see 0.9 GHz panel of Fig.\ref{fig:radio_lc}). In subsequent observations, its flux density declines steadily for $>1000$ days. Although found within the GW190814 localisation region, the transient is unrelated as its potential host’s redshift (see Section \ref{sec:host_galaxy}) places it outside the gravitational-wave distance range, implying a chance sky coincidence. 

Motivated by this discovery and evolution, we searched for multi-wavelength counterparts. We cross-matched the coordinates of \source\ with the Transient Name Server\footnote{\url{https://www.wis-tns.org}} and found no association within $1\arcmin$. Likewise, no known active galactic nuclei (AGN) or quasars were identified along the line of sight within $5\arcmin$ using the Million Quasars (Milliquas) catalogue \citep{2023OJAp....6E..49F}. We also found no matches in archival X-ray catalogues hosted by NASA\footnote{\url{https://heasarc.gsfc.nasa.gov/cgi-bin/W3Browse/w3browse.pl}} within $15\arcsec$ of the source. 

We queried the GRBWeb catalogue by Paul Coppin\footnote{\url{https://user-web.icecube.wisc.edu/~grbweb_public/index.html}} \citep{PhysRevD.102.103014} for GRBs occurring before our earliest radio detection (2022 February 4) whose localisation regions overlap the position of \source. We identified 54 such GRBs over the past $\sim 30$\,years. Given the reported localisation areas and the average all-sky GRB detection rate, the chance-coincidence probability for any single burst to overlap our position exceeds $50\%$; none therefore constitutes a compelling counterpart (see Appendix~\ref{app:mw} for details). We also used  \texttt{SWIFTBAT\_python}\footnote{\url{https://github.com/lanl/swiftbat_python}}, and found that the source position was in the BAT field of view $\sim 65\%$ of the time (for a 1000\,s cadence search) during the two years preceding our first radio detection. No counterpart is seen immediately before the radio discovery, although gaps in coverage leave open the possibility that a prompt event was missed.

\source\ lies $4.1\arcsec$ from the galaxy 2dFGRS~TGS143Z140. If associated with this galaxy (Section~\ref{ss_host_disc}), the peak specific luminosity would be $\sim 10^{30}\,\rm erg\,s^{-1}\,Hz^{-1}$, making it more luminous and longer lived than all known supernovae \citep{2021ApJ...908...75B}.

\section{Observations and Data Processing} \label{sec:data}

To characterise \source\ we conducted a comprehensive multi-wavelength observing campaign. In this Section, we describe the observational setup and reduction pipelines used for each facility.

\subsection{Radio Observations}

Following our discovery with ASKAP, we conducted targeted follow-up with the Australia Telescope Compact Array (ATCA), upgraded Giant Metrewave Radio Telescope (uGMRT), MeerKAT and Parkes. All radio data used in this work are tabulated in Appendix~\ref{app:radio} and shown in Fig.~\ref{fig:radio_lc}, including the early 0.9\,GHz ASKAP non-detections. Our best source position, derived from ATCA and GMRT observations, is \((\text{R.A.}_{\text{J2000}},\text{Decl.}_{\text{J2000}})=00{:}55{:}12.21,\,-25{:}58{:}34.9\), with an uncertainty of \(0.1^{\prime\prime}\).

\subsubsection{ASKAP}
\label{ss_ASKAP_obs}
ASKAP~J0055$-$2558 was detected in the VAST survey \citep{2021PASA...38...54M}, which observes at 888\,MHz with a roughly bimonthly cadence. Typical 12–15\,min integrations reached $\sim$0.3\,mJy rms.

Short-timescale (minutes) variability and polarisation provide powerful diagnostics of scintillation or coherent bursts (whereas weakly polarised emission favours synchrotron). To test these possibilities, we also analysed the 943\,MHz ASKAP observations on 2024-01-02 with \textsc{DStools}\footnote{\url{https://github.com/askap-vast/dstools}} \citep{Pritchard2025a}. After subtracting field sources with \textsc{WSclean}, we formed full-Stokes time–frequency (dynamic) spectra at the source by averaging calibrated visibilities over all antenna pairs, produced 50\,s–binned light curves, and computed modulation indices.

\subsubsection{ATCA}

We observed \source\ with ATCA for nine epochs between May 4, 2023, and March 20, 2025 (Project code C3363; PI: T. Murphy), measuring the flux density at a range of frequencies to determine its evolution. All our observations used 2048 MHz-wide bands centred on 2.1, 5.5, and 9.0\,GHz. For the L-band ATCA observations, we also produced sub-band images: the 2048-MHz band was split into four 512-MHz sub-bands centred at 1.332, 1.844, 2.100, and 2.356 GHz, and each sub-band was imaged independently. In addition to the independently imaged sub-bands, we also present the flux density at 2.1 GHz, which corresponds to the full 2048-MHz band. We considered a source detected if its measured flux exceeded five times the local image rms (5$\sigma$); measurements below this threshold were treated as non-detections and reported as upper limits. We reduced the visibility data using standard routines in \texttt{MIRIAD} \citep{1995ASPC...77..433S}. We used a combination of manual and automatic radio frequency interference (RFI) flagging before calibration. We used PKS~B1934$-$63 to determine the bandpass response and calibrate the flux density scale for all epochs and frequency bands. We used PKS~0023$-$26 to calibrate the gains for all frequency bands. After calibration, we inverted the visibilities using a robust weighting of 0.5 and then used the CLEAN algorithm \citep{1980A&A....89..377C} with 3000 iterations to the target source field. We also searched 2100\,MHz ATCA images on 2023-08-27 for short-timescale variability and polarised emission as described in Section~\ref{ss_ASKAP_obs}.

\begin{figure*}[t]
  \centering
  \begin{minipage}[t]{1\textwidth}
    \vspace{0pt} 
    \centering
    \includegraphics[width=\linewidth]{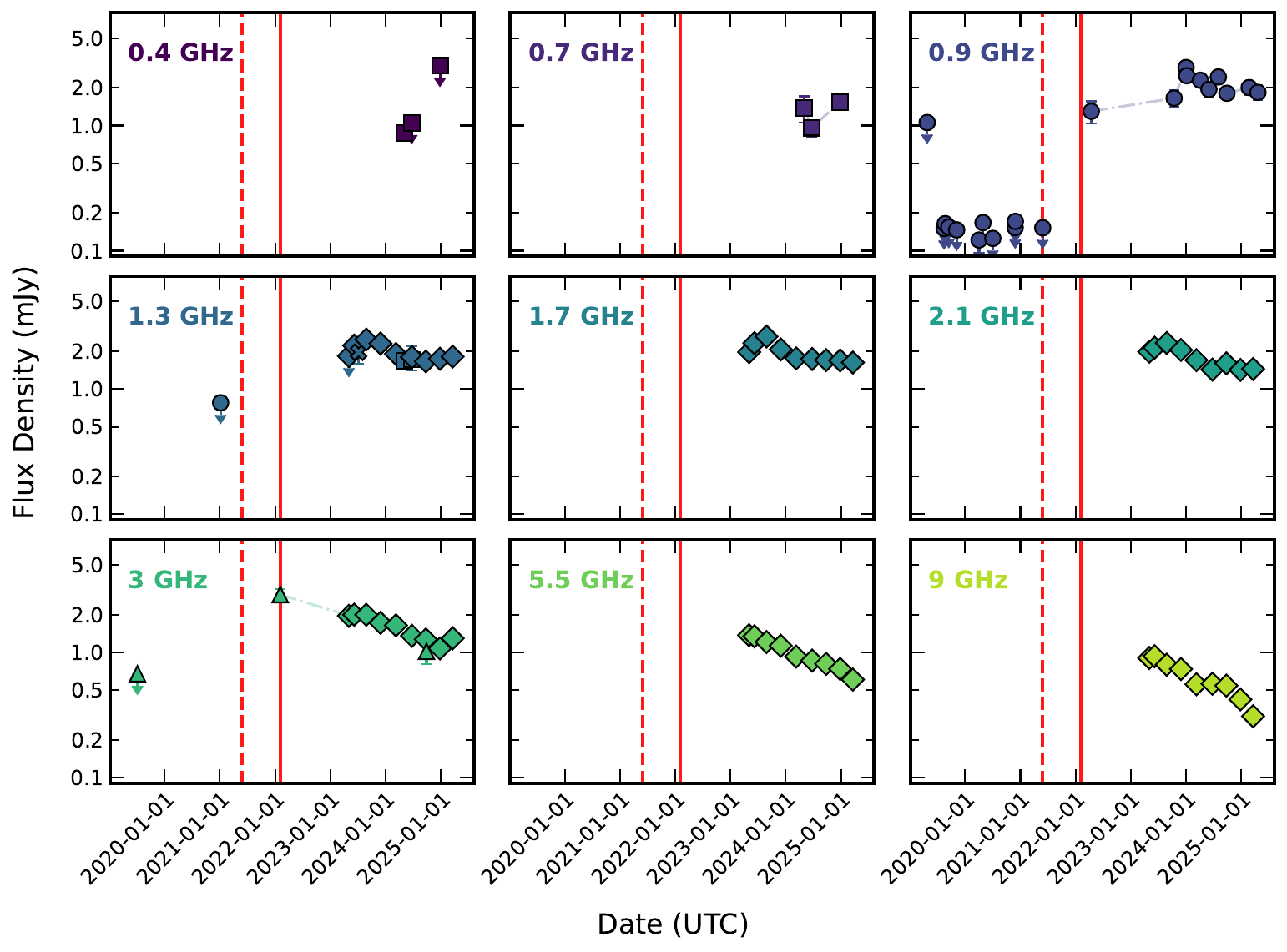}
  \end{minipage}\hfill
  
  \caption{Radio light curves of ASKAP J0055–2558. The data span frequencies from 0.4 to 9 GHz, obtained with ASKAP ("$\circ$"), ATCA ("$\diamond$"), GMRT ("$\square$"), VLA ("$\triangle$") and MeerKAT ("$\boldsymbol{\times}$"). Different frequency bands are represented by distinct colours. 5$\sigma$ upper limits are indicated by downward arrows. The dashed red vertical line marks the last non-detection prior to the onset of detections, while the solid red vertical line marks the first radio detection.}
  \label{fig:radio_lc}
\end{figure*}

\subsubsection{uGMRT}

We conducted observations with the
uGMRT in 2024 May, June and December (Project code 46$\_$054). All our observations used 400 MHz-wide bands centred on  0.67 and 1.27\,GHz, and 200 MHz-wide bands centred on 0.4\,GHz. We reduced the visibility data using the Common Astronomy Software Applications (\texttt{CASA}) package \citep{2007ASPC..376..127M}. We used a combination of manual and automatic RFI flagging before calibration. 3C48 was used as primary and bandpass calibrator (flux set by \citet{2017ApJS..230....7P} scale), while 0025$-$260 was used for gain calibration. Imaging used w-projection convolution to correct wide-field non-coplanar baselines \citep{2008ISTSP...2..647C} and Briggs robust=0 weighting, yielding higher angular resolution than natural weighting at the cost of slightly increased noise. At uGMRT low frequencies, ionospheric delays induce phase errors \citep{2019MNRAS.483.1184M}, mitigated via self-calibration—iteratively solving on compact, bright sources and re-imaging to refine the model \citep{1984ARA&A..22...97P}. We ran four phase-only self-cal cycles (solution intervals 4, 2, 1, and 0.5 min) followed by four amplitude-and-phase cycles (8, 4, 2, and 1 min) on the target visibilities. The phase calibrator set below the horizon after the initial scans for band 5, resulting in significant phase errors in the 26 December 2024 observation. Consequently, this data has been excluded from the analysis.


\subsubsection{MeerKAT}

We observed with the MeerKAT array \citep{Jonas} in L-band (856–1712\,MHz) on 2023 July 7, using the c856M4k mode with $\sim$770\,MHz of usable bandwidth and 8\,s integrations. Full-polarisation data from the on-source beam were recorded (but not searched in real time) with the PTUSE backend of the MeerTime pulsar timing project \citep{meertime2020}, in search mode with 38.28\,$\mu$s sampling and stored in \texttt{psrfits} format. We carried out periodicity searches with \texttt{pulsar\_miner}\footnote{\url{https://github.com/alex88ridolfi/PULSAR_MINER}} \citep{2021MNRAS.504.1407R}, a wrapper for PRESTO \citep{2011ascl.soft07017R}, dividing the data into 5\,min segments. We searched dispersion measures up to 50\,pc\,cm$^{-3}$, twice the maximum Galactic DM along this sightline predicted by the YMW16 Galactic electron-density model \citep{2017ApJ...835...29Y}, and periods from 1\,ms to 10\,s, including Fourier-domain acceleration searches with $z_{\max}=\pm200$. No convincing pulsar-like candidates were detected with $\mathrm{S/N}>8$.

\subsubsection{Parkes}

We monitored \source\ for short-duration bursts with the Ultra-Wideband Low (UWL) receiver on Murriyang, the 64\,m Parkes telescope, covering 704–4032\,MHz continuously \citep{Hobbs2019ultrawide}. A 1\,hr observation on 2023 November 18 was conducted in single-polarisation pulsar search mode, with 250\,kHz spectral resolution (13312 channels), 64\,$\mu$s sampling, and data recorded in 2-bit \texttt{psrfits} search format \citep{Hotan04}. We searched for bursts using a subbanded pipeline \citep{2022MNRAS.512.3400K} that integrates \textsc{Heimdall} \citep{Barsdell12} and \textsc{FETCH} \citep{Aggarwal20}. The search covered 0–1000\,pc\,cm$^{-3}$ dispersion measures (step 0.1) and boxcar widths up to 2\,min, dividing the band into overlapping subbands (64–3328\,MHz) to avoid edge losses. All \textsc{FETCH} candidates were visually inspected, and no events with $\mathrm{S/N}>7$ were found.

\subsection{Optical/IR Observations}

We obtained optical and near-infrared imaging with Magellan and Keck telescopes, and spectroscopy with Keck and the ANU 2.3-m telescope, to characterise the candidate host galaxy morphology and constrain any excess emission at the transient position.

\subsubsection{Magellan Imaging}
\label{Magellan_imaging}

We observed \source\ with the Low Dispersion Survey Spectrograph (LDSS-3) on the 6.5-m Magellan Clay Telescope at Las Campanas Observatory on 2024 November 26 \citep[e.g.,][]{2008SPIE.7014E..0AO}. Eighteen 300s images were taken using the Sloan $g^\prime$ filter, for a total exposure of 1.5 hr, under $\approx 0.7^{\prime\prime}$ seeing conditions. Images were flat-fielded, sky-subtracted, and combined using a custom data reduction pipeline based in part on \texttt{ccdproc} \citep{ccdproc}, and cosmic rays were removed with the Laplacian edge detection algorithm implemented in \texttt{astroscrappy} \citep{van_dokkum2001,astroscrappy}. Astrometric and photometric calibration were performed using standard data reduction pipelines, including \texttt{SExtractor} \citep{Bertin1996} for aperture photometry calibrated to the Pan-STARRS1 system \citep{2012ApJ...750...99T,2016arXiv161205560C}.

\subsubsection{ANU 2.3m WiFeS IFU spectroscopy}

On 2023 July 12, we obtained spectra around the position of the radio source with the ANU 2.3m telescope at Siding Spring Observatory \citep{2013JAHH...16....2M}. We used the Wide Field Spectrograph \citep[WiFeS;][]{Dopita:2007}, which provides a $25^{\prime\prime}\times 38^{\prime\prime}$ field of view, enabling integral field unit (IFU) spectroscopy of both the radio source position and the candidate host galaxy in a single pointing. We obtained $R=3000$ resolution spectra using the R3000 and B3000 gratings, for the red and blue arms of the spectrograph and the RT560 dichroic covering a wavelength range of 3500--9000\,\AA. To obtain a better signal-to-noise ratio, we use the Nod \& Shuffle (N\&S) mode. Following the configuration used in \citet{Moller:2025}, we observed three N\&S $1200$\,s exposures ($600$\,s on target, $600$\,s on sky). The ANU 2.3m IFU cube only detected the main component of the potential host galaxy.

\subsubsection{Keck KCWI/KCRM IFU spectroscopy}

On 2023 August 17, we observed the candidate host galaxy and surrounding region with the Keck Cosmic Web Imager (KCWI; blue arm) and the Keck Cosmic Reionisation Mapper (KCRM; upgraded red arm) on the 10\,m Keck-2 telescope, as an alternate target during poor weather for an unrelated program (W377; PI Deller). We obtained $2\times290$\,s exposures with KCWI and $2\times300$\,s exposures with KCRM. We used the BL and RL gratings for KCWI and KCRM, respectively, and the medium slicer for both instruments, providing a $16.5^{\prime\prime}\times 20.4^{\prime\prime}$ field of view. We calibrated and reduced the data using standard \texttt{PypeIt} \citep{2020JOSS....5.2308P} routines and formed spectral cubes for each dataset covering wavelengths 3420--5590$\AA$ using KCWI and 6502--10360$\AA$ using KCRM. We extracted the spectra of the galaxy nucleus as well as the surrounding bright components using \texttt{QFitsView} and analysed the reduced data with the \texttt{LIME} package \citep[][]{2024A&A...688A..69F}. 

\subsubsection{Keck MOSFIRE}
\label{ss:IR}

Near-infrared J-band ($1.2\,\mu$m) imaging for \source\ on 2024 September 14 was acquired using the Multi-Object Spectrometer For InfraRed Exploration (MOSFIRE; \citealt{2010SPIE.7735E..1EM,2012SPIE.8446E..0JM}) mounted on the 10m Keck 1 telescope. We obtained 200 $\times$ 10\,s images for a total of 2000\,s using a ``box5" dithering pattern under moderate seeing conditions (average $\sim$0.8 arcsec seeing FWHM measured from the stacked image). The data were reduced and median stacked using IRAF \citep{Tody86} and SWARP \citep{bertin2012}, and sources were detected using SExtractor \citep{Bertin1996}.

\subsection{X-ray Observations}

\subsubsection{SWIFT}
\label{ss_xray}

We observed ASKAPJ0055$-$2558 with the X-ray Telescope (XRT; \citealp{Burrows:2005}) on board \textit{Swift} on 2024 August 8 (916 days post first radio-detection) for 3.6\,ks (ObsID 00016752001; PI: Gulati). We processed the data with \textsc{xrtpipeline}, following standard guidelines (e.g., \citealp{Evans:2009}). We detect zero counts in the 0.2--12, 0.2--2, or 2--12\,keV images which gives us an upper count rate limit of $6.4\times10^{-4}\,\mathrm{ct\,s^{-1}}$ in any band, corresponding to 90\% upper limits on the Galactic-absorbed flux of $2.9\times10^{-14}$, $1.1\times10^{-14}$, and $1.8\times10^{-14}\,\mathrm{erg\,cm^{-2}\,s^{-1}}$. Fluxes were computed with \texttt{PIMMS}\footnote{\url{https://heasarc.gsfc.nasa.gov/docs/software/tools/pimms.html}} assuming a power-law spectrum with photon index $\Gamma=1.7$ and Galactic line-of-sight absorption $N_{\rm H}=2\times10^{20}\,\mathrm{cm^{-2}}$, appropriate given the transient’s location outside the main body of the potential host galaxy \footnote{Calculated with the NASA HEASARC $N_{\rm H}$ tool: \url{https://heasarc.gsfc.nasa.gov/cgi-bin/Tools/w3nh/w3nh.pl}}. This choice is motivated by late-time, optically thin synchrotron emission (for $p\!\sim\!2.2$–2.4, $\Gamma\!\approx\!(p+1)/2\!\sim\!1.6$–1.7; \citealp{2002ApJ...568..820G}) and is consistent with typical \textit{Swift}/XRT afterglow spectra \citep{Evans:2009}. To illustrate sensitivity to the assumed spectrum, adopting $\Gamma$ in the range 1.5–2.1 changes the flux limits in the 0.2–12\,keV band by $\lesssim$20\% for the same $N_{\rm H}$; we therefore regard our non-detection and derived constraints as robust.

\section{Transient Properties}
\label{s_transient_identification}

In this section, we present
the observational constraints from our multi-wavelength campaign and the radio characterisation of \source. Firstly, we summarise the non-detections that rule out several common source classes.

The source has no infrared counterpart in NIR J-band to 23\,mag (Section~\ref{ss_IR_hg}), inconsistent with a typical background AGN, no optical transients reported at this location. Despite persistent radio emission, near-simultaneous X-ray follow-up yields no detection (Section~\ref{ss_xray}); with 90\% upper limits on the Galactic-absorbed flux reaching the few nJy level. L-band searches with MeerKAT and Parkes UWL reveal no pulsed emission, excluding Galactic pulsars and $\gamma$-ray binaries. From the 5.5\,GHz ATCA epoch on 2023 June 8, we constrain circular polarisation to $V/I<5.6\%$ (5$\sigma$; $S_V=0.08$\,mJy, $S_I=1.39$\,mJy) and find no significant signal in Stokes $Q$, $U$, or $V$. These limits rule out emission dominated by strongly circularly polarised coherent processes (e.g. pulsars or flare-stars) and are consistent with unpolarised synchrotron emission. An extreme scattering event is also unlikely as even though large plasma lenses can, in principle, yield $>20\times$ magnification, the associated fractional bandwidth is typically narrow unless a contrived lens is invoked \citep[e.g.][]{2016ApJ...817..176T}; while a broadband de-magnification followed by a brief enhancement could mimic early non-detections and a trailing horn \citep{1987Natur.326..675F,2016Sci...351..354B}, the observed years-long, broadband decline, strong at 5.5 and 9\,GHz with little variation below 2\,GHz, rules out an ESE origin.

In the following subsections, we present the radio spectral, temporal evolution, and variability analyses.

\subsection{Radio variability}
\label{ss_ISS}

We searched for variability on timescales of minutes to months. There was no minute-scale variability within the 15-min scans (with modulation indices $0.10\pm0.10$ (ATCA) and $0.3\pm0.3$ (ASKAP). The expected timescale of diffractive interstellar scintillation is 6.8 and 18.9\,min, respectively; hence, this can be ruled out for \source\ . On longer timescales, however, we do detect variability at 0.9\,GHz (ASKAP) consistent with refractive ISS (RISS), which yields only a loose size limit, $\theta_s \lesssim 0.5^{\prime\prime}$. Because this is not constraining, we instead predict the RISS contribution and fold it into our flux uncertainties.

Assuming ASKAP~J0055$-$2558 is a compact synchrotron source, we expect RISS to modulate its flux density. We used the NE2001 Galactic electron-density model \citep{2002astro.ph..7156C} to evaluate the line-of-sight scattering measure at the source Galactic coordinates. Adopting the relations in \citet{1998MNRAS.294..307W} to convert the scattering measure to a transition frequency, we obtain 
$\nu_0=7.46$\,GHz, i.e. the frequency separating the strong and weak scattering regimes. Following \citet{1998MNRAS.294..307W} for extragalactic sources, we compute the scattering-disk angular size $\theta_r$, which sets the angular scale below which a source remains in the strong-scintillation regime at a given observing frequency (Table~\ref{tab:iss_params}).

We estimate the intrinsic source size via a preliminary equipartition analysis (Section~\ref{ss_eqipartition_analysis}, RISS errors not included, see Appendix \ref{app:radio} Fig. \ref{fig:SBPL_params} top panel for fitted spectral parameters). For all assumed geometries other than the $5^\circ$ angle geometry ($\theta_s=1247\mu as$), the mean source size $\theta_s$ is smaller than $\theta_r$ for $\nu\lesssim2.3$\,GHz, whereas at 5.5\,GHz the mean $\theta_s$ exceeds $\theta_r$ in every geometry. Accordingly, we report modulation indices for two representative sizes: $230\,\mu$as (less-collimated geometries) and $1247\,\mu$as (the $5^\circ$ cone). The RISS modulation index $m$ from \citet{1998MNRAS.294..307W} is
\begin{equation}
m =
\begin{cases}
\left(\dfrac{\nu}{\nu_0}\right)^{17/30}, & \theta_s \le \theta_r,\\
\left(\dfrac{\nu}{\nu_0}\right)^{17/30}\left(\dfrac{\theta_r}{\theta_s}\right)^{7/6}, & \theta_s > \theta_r,
\end{cases}
\end{equation}
where
\begin{equation}
\theta_r=\theta_{F0}\left(\dfrac{\nu}{\nu_0}\right)^{-11/5},
\end{equation}
and $\theta_{F0}$ is the angular size of the first Fresnel zone.

\begin{deluxetable}{ccccc}
\tablecaption{Estimated RISS Parameters for ASKAP~J0055$-$2558 \label{tab:iss_params}}
\tablehead{
\colhead{$\nu$} & 
\colhead{$m_p$} & 
\colhead{$\theta_{F0}$} & 
\colhead{$\theta_r$} & 
\colhead{$m$} \\
\colhead{(GHz)} & 
\colhead{} & 
\colhead{($\mu$as)} & 
\colhead{($\mu$as)} & 
\colhead{}
}
\startdata
0.4 & 0.19 & 2076 & 1296346 & 0.19 \\
0.7 & 0.26 & 790  & 144024  & 0.26 \\
0.9 & 0.30 & 455  & 47720   & 0.30 \\
1.3 & 0.37 & 202  & 9434    & 0.37 \\
1.7 & 0.43 & 112  & 2899    & 0.43 \\
2.1 & 0.49 & 70   & 1138    & 0.49 \\
2.3 & 0.51 & 58   & 772     & 0.51/0.30 \\
3.0 & 0.60 & 32   & 237     & 0.60/0.09 \\
5.5 & 0.84 & 8    & 16      & 0.04/0.005 \\
9.0 & 0.76 & 3    & 2       & 0.005/0.001 \\
\enddata
\tablecomments{Parameters include the point-source modulation index ($m_p$), Fresnel-zone size ($\theta_{F0}$), scattering-disk size ($\theta_r$), and the final modulation index ($m$) after accounting for source size. For $\nu \ge 3$\,GHz, two $m$ values correspond to $\theta_s=230$ and $1247\,\mu$as, respectively. Values are based on NE2001 and are used to compute additional flux uncertainties from RISS.}
\vspace{-1cm}
\end{deluxetable}

Before spectral fitting, we add a RISS term in quadrature to each flux-density uncertainty, equal to $m(\nu,\theta_s)\times S_\nu$. This contributes 19\% at 0.4\,GHz, 51\% at 2.3\,GHz, and $\sim$1\% at 9\,GHz (weak regime). Because the $5^\circ$ geometry yields smaller $m$ at $\nu\!\ge\!3$\,GHz, we conservatively adopt the larger uncertainties from the less-collimated case for all spectral fits. All RISS contributions used in the modelling are reported in Table~\ref{Table:Radio_obs}.

\subsection{Radio Spectral Evolution}
\label{ss_spectra_temp_modelling}

\begin{figure*}[t]
  \centering
  \begin{minipage}[t]{1\textwidth}
    \vspace{0pt} 
    \centering
    \includegraphics[width=\linewidth]{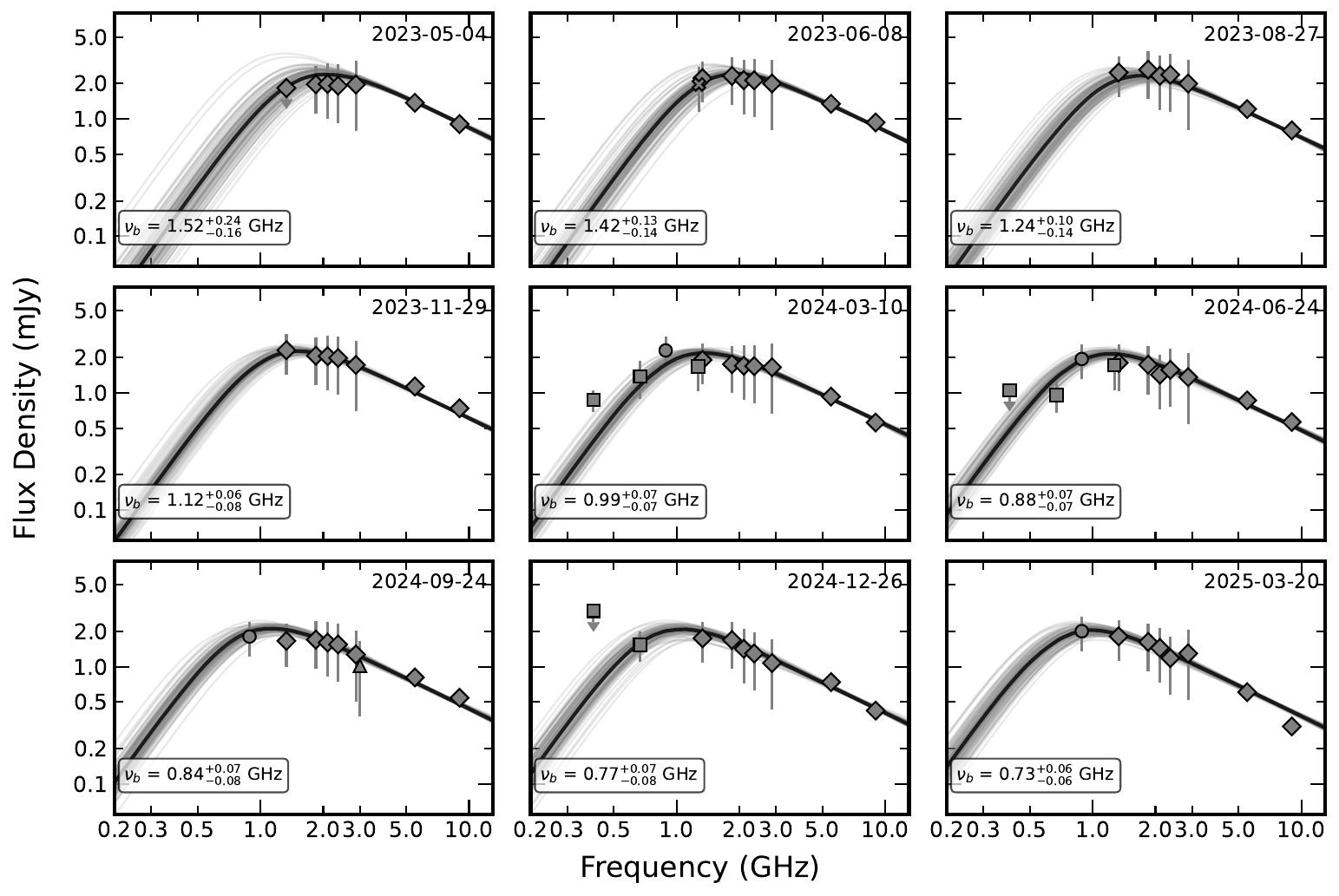}
  \end{minipage}
  \caption{Radio spectral evolution of \source\ over nine ATCA epochs from 2023 to 2025. The errors on datapoints include both measurement and Interstellar Scintillation (ISS) errors. Dark solid curves show the best-fitting time-evolving SBPL model obtained from a joint fit to all epochs, with the peak at each epoch set by the self-absorption break; faint thin curves show 50 random draws from the MCMC posterior.}
  \label{fig:radio_spectra}
\end{figure*}

\source\ exhibits a peaked radio spectrum in our broadband radio data. Such spectra can arise from two generic mechanisms: (i) synchrotron self-absorption (SSA), in which the low-frequency turnover is intrinsic to a compact, recently activated (age $\lesssim 10^{5}$ yr) jet, lobe, or blast wave \citep{chevalier_synchrotron_1998,1997AJ....113..148O,1998A&A...337...69O}; and (ii) free–free absorption (FFA), in which emission is attenuated by dense ionized plasma along the line of sight \citep{weiler_radio_2002}—either homogeneous or inhomogeneous \citep[e.g. in Compact Steep Spectrum, CSS, and Gigahertz Peaked Spectrum, GPS, sources;][]{1984AJ.....89....5V}. Slower outflows (e.g. SN ejecta or cocoons) can also show SSA/FFA, and both processes may operate simultaneously (as seen in radio SNe). In epochs with adequate low-frequency coverage, we therefore fit the optically thick index as a free parameter, following \citet{2022MNRAS.512.5358R} (RISS errors not included), to distinguish among SSA, homogeneous FFA, and inhomogeneous FFA. Only SSA and inhomogeneous FFA reproduce the observed rising, optically thick spectrum of \source\ below the turnover (mean $\alpha_1$ of \(1.2\pm0.4\)); homogeneous FFA yields an exponential cutoff inconsistent with the data.

While our low-frequency coverage is not sufficient to fully disentangle SSA from FFA, a GPS/CSS interpretation is specifically disfavored due to the absence of an IR/optical counterpart at the radio position (unusual for GPS/CSS), a steep post-peak temporal decline that is faster than typical GPS/CSS evolution \citep{2016AN....337....9O,2022MNRAS.512.5358R}, and a low chance alignment probability with the potential host galaxy (Section~\ref{sec:host_galaxy}). With homogeneous FFA excluded by the spectral shape, and given that our low-frequency coverage cannot separately constrain any inhomogeneous FFA contribution, we adopt an SSA-only baseline model for a compact, recently activated jet associated with the host galaxy. 

Assuming an SSA origin, we performed a preliminary fit to each epoch’s spectrum, including scintillation uncertainties, using a smoothly broken power-law (SBPL) model expressed as:

\begin{equation}
    S(\nu) = A \left[ \left( \frac{\nu}{\nu_b} \right)^{-s\alpha_1} + \left( \frac{\nu}{\nu_b} \right)^{-s\alpha_2} \right]^{-1/s}
\vspace{-0.1cm}
\end{equation}
Where $S(\nu)$ is the flux density at frequency $\nu$, $\nu_b$ is the break (or peak) frequency, $A$ is a normalization constant (corresponding to the peak flux density), $\alpha_1$ and $\alpha_2$ are the spectral indices below and above the break, respectively, and $s$ is a smoothness parameter. We fixed $s = 1$ to ensure a gradual transition between spectral regimes. This choice is physically motivated, as the aggregate optically thin spectral index $\alpha \simeq -0.9$ (for frequencies $>2$~GHz) implies an electron index $p \simeq 2.7$. Using the empirical relations $s = 1.47 - 0.21\,p$ (ISM) and $s = 1.25 - 0.18\,p$ (wind) from \citet{2002ApJ...568..820G}, we obtain $s \approx 0.7$--$0.9$, that is, an order-unity smoothness consistent with our adopted value of $s = 1$. Following \citet{2002ApJ...568..820G,mooley_late-time_2022}, we fixed the value of $\alpha_1=5/2$ consistent with the optically-thick spectral index of a self-absorbed synchrotron spectra i.e. $\nu_m<\nu_a$, where $\nu_m$ is the characteristic frequency of electrons at the minimum Lorentz factor and $\nu_a$ is the synchrotron self-absorption frequency below which the emission is optically thick. We implemented the model in \texttt{Python}, and spectral fitting was performed using the \texttt{emcee} package \citep{2013PASP..125..306F}, an affine-invariant Markov Chain Monte Carlo (MCMC) ensemble sampler.

The fitted peak flux density and frequency are either constant or declining and the optically thin index remains at $\alpha_{2} \approx -0.9$ (see Appendix~\ref{app:radio} Fig. \ref{fig:SBPL_params} middle panel). Apart from the three GMRT epochs with low-frequency coverage, most epochs are sparsely sampled below the spectral peak, leaving peak parameters poorly constrained or even unconstrained for epochs 2023-05-04 and 2023-08-27. This is compounded by large flux uncertainties near the peak, where interstellar scintillation is strong (Section~\ref{ss_ISS}). To mitigate these issues, we fit all epochs jointly with weak, uniform priors and shared parameters. This time-dependent fit (including ISS error inflation; Section~\ref{ss_ISS}) constrains the evolution of peak frequency and amplitude. The spectrum is modelled as a smoothly broken power law with fixed optically thick index $\alpha_1$, global (epoch-independent) $\alpha_2$ and smoothness $s$, break frequency and amplitude evolving with time as a power law (index positive, negative, or zero).

Specifically,
\vspace{-0.3cm}

\begin{equation}
\begin{aligned}
F(\nu,t) &= A(t)\,\left[\left(\frac{\nu}{\nu_b(t)}\right)^{-s\alpha_1}
           + \left(\frac{\nu}{\nu_b(t)}\right)^{-s\alpha_2}\right]^{-\frac{1}{s}},\\[4pt]
\nu_b(t) &= \nu_{b,0}\left(\frac{t}{t_0}\right)^{m},A(t) = A_0\left(\frac{t}{t_0}\right)^{b} \\[4pt].
\end{aligned}
\end{equation}

\vspace{-0.8cm}

The amplitude $A(t)$ evolves as a single power law in time with slope $b$, and is normalised such that $A(t_0) = A_0$ at the reference epoch $t_0$. The break frequency $\nu_b(t)$ evolves as a power law with index $m$ from the reference value $\nu_{b,0}$. Since limited sub-GHz coverage leaves \(\alpha_1\) weakly constrained, we fix \(\alpha_1=5/2\) (SSA-like) and $s=1$ (spectral low-energy slope and smoothness), while fitting the parameters $\{A_0,\nu_{b,0},m,b,\alpha_2\}$. The priors were allowed to vary uniformly within the following ranges: 
$A_0 \in [0, 1000]\,\mathrm{mJy}$; 
$\nu_{b,0} \in [0.1, 100]\,\mathrm{GHz}$; 
$m \in [-5, 5]$; 
$b \in [-5, 5]$; 
$\alpha_2 \in [-5, 0]$.

\begin{figure}[t]
    \centering
    \includegraphics[width=\linewidth]{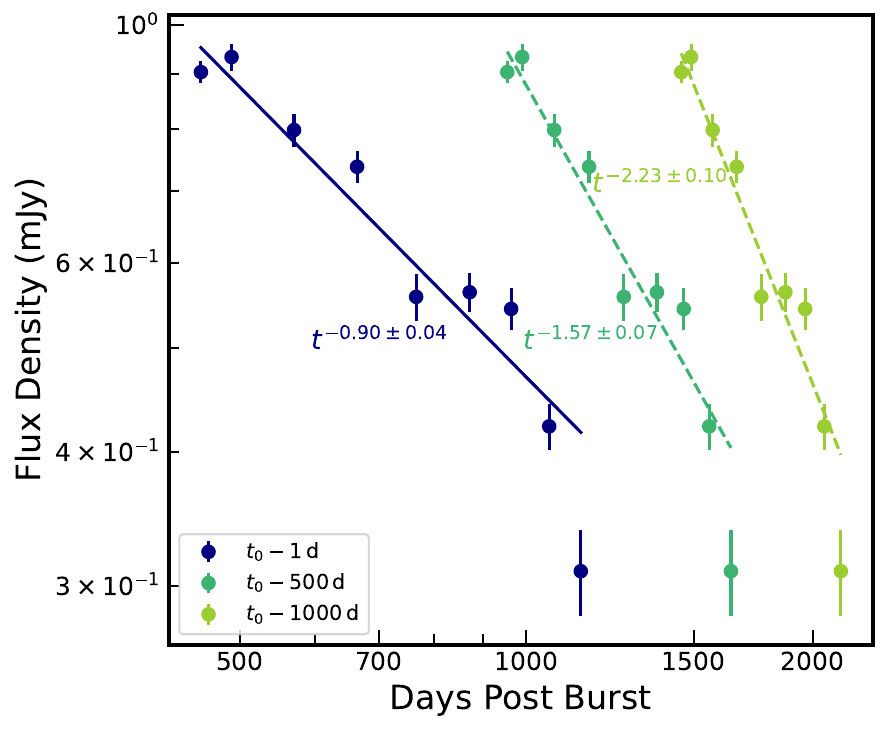}
   
    \caption{The 9 GHz light curve of \source, where the temporal index is derived by shifting the reference epoch of the first radio detection ($t_o$ = 2022-02-04) by +1, +500, and +1000 days. The blue, teal, and green lines show the corresponding power-law fits.}
    \label{Figure:temp_fit}
\end{figure}

The observed and modelled spectra are shown in Fig.~\ref{fig:radio_spectra} (see Appendix \ref{app:radio} Fig. \ref{fig:SBPL_params} bottom panel for fitted spectral parameters). The optically thin spectral index from the combined fit is $\alpha_{\text{2}}= -0.86\pm0.04$, implying an electron energy distribution with a power-law index of $p=2.72\pm0.08$, consistent with synchrotron emission \citep{2002ApJ...568..820G}. From the joint fit to all epochs, we find a reference break frequency of $\nu_{b,0}=0.97^{+0.07}_{-0.07}~\mathrm{GHz}$ with temporal evolution $m=-0.80^{+0.22}_{-0.21}$ (i.e., $\nu_b \propto t^{\,m}$). The amplitude slope is $b=-0.18^{+0.18}_{-0.19}$. Extrapolating our fit to higher optical and X-ray frequencies for the epochs with near-simultaneous observations, we find that the upper limits are consistent with the non-detections.

We also fit the datapoints at 9\,GHz with power law $F_\nu\propto t^b$ (see Fig. \ref{Figure:temp_fit}), to estimate the transient's temporal evolution. Since the time of explosion is not known, we try fiducial values of 1, 500 and 1000 days for the age of the transient at the time of first radio detection at epoch 2022-02-04 and we find $F_{9 GHz}\propto t^{-0.9\pm0.04}$, $F_{9\rm{GHz}}\propto t^{-1.57\pm0.07}$ and $F_{9 GHz}\propto t^{-2.23\pm0.10}$ respectively. The final radio data point deviates from the single power-law fit; this could be a measurement artefact, or genuine late-time steepening, also seen in orphan the afterglow candidates FIRST J141918.9+394036 \citep[][]{2022ApJ...924...16M}, and jetted TDE Swift J1644+57 \citep[][]{2018ApJ...854...86E}. Temporal evolution is shallower for lower frequencies, which should be expected depending on where the observation frequency lies with respect to the characteristic synchrotron peak frequency \citep{2002ApJ...568..820G}.

\section{Potential Host Galaxy}
\label{sec:host_galaxy}

\begin{figure*}[t]
    \centering
    \includegraphics[width=\textwidth]{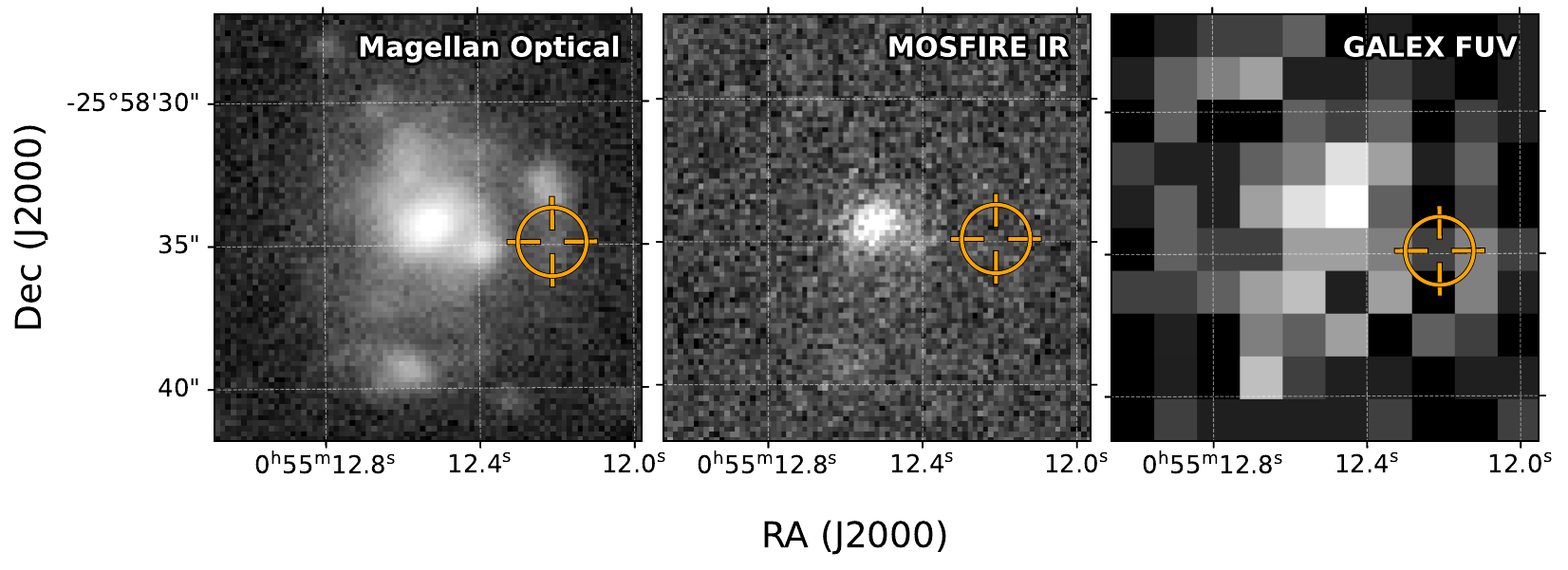}
    
    \vspace{0.5cm} 

    \vspace{-0.7cm}\includegraphics[width=\textwidth]{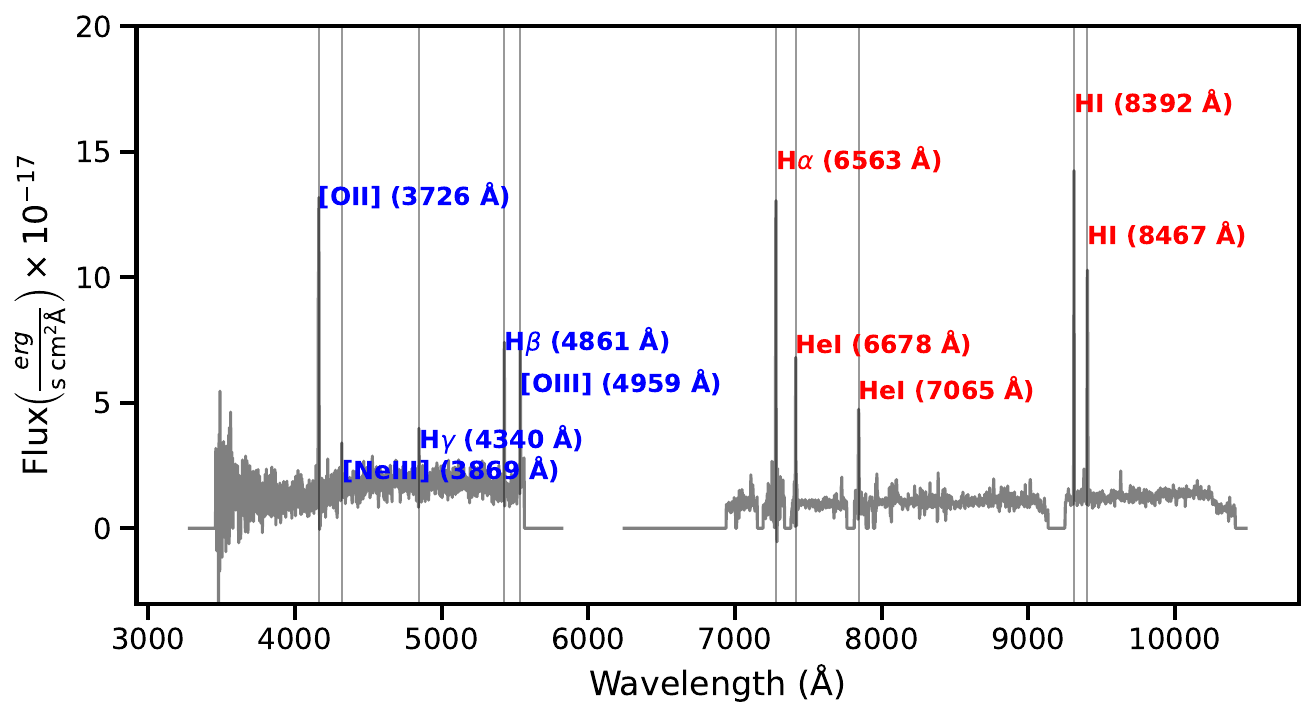}
    
    \caption{Top: Magellan ($5\sigma$ AB limiting magnitude 28.0 in g'-band), MOSFIRE (23.5 in J-band), and GALEX FUV (22.6) images of the transient position overlaid by the aggregate source position  derived from ATCA and GMRT observations (the circle around the position corresponds to the aggregate PSF with a radius of 1.2 arcsec). Bottom: Spectra obtained with Keck KCWI and KCRM for the galaxy nucleus, with emission lines fitted using the \texttt{LIME} package. The signal-to-noise ratio, as defined by \citet{1994A&A...287..676R} for each of the fitted lines, is above 5$\sigma$. The fitting results are given in the Appendix \ref{app:mw} Fig. \ref{fig:optical-fits}.}
    \label{fig:combined_overlay_spectra}
\end{figure*}

\source\ is offset \(4.1^{\prime\prime}\) from the galaxy 2dFGRS TGS143Z140 (see Fig. \ref{fig:combined_overlay_spectra}), which has a confirmed spectroscopic redshift of \(z = 0.116\) \citep{2001MNRAS.328.1039C}. Using the angular diameter distance of 436\,Mpc, this corresponds to a projected physical offset of approximately 8.7\,kpc. The probability of a chance alignment between \source\ and 2dFGRS TGS143Z140 is only 1.87\% \citep[computed using the formalism of][see Appendix \ref{app:hg_chance_allignment}]{2002AJ....123.1111B}, supporting a likely association. The galaxy is clearly detected in the DESI Legacy Surveys \textit{DR10} \citep{2019AJ....157..168D}, with catalogued z, i, r, and g magnitudes of 19.27, 19.33, 19.58, and 20.01 mag, respectively. It is also UV-luminous, identified as GALEXASC J005512.43–255832.9, with far-UV and near-UV magnitudes of 21.53 and 21.06 mag \citep[DR6;][]{2005ApJ...619L...1M}. 

\subsection{Spectroscopy}

We extracted the galaxy spectrum from the ANU 2.3 m spectral cube, revealing $H\alpha$ and oxygen emission lines. Using the software {\sc MARZ} \citep{Hinton:2016}, we found that the spectrum is consistent with a galaxy at redshift $0.116$. Using the more sensitive Keck spectra, the fitted redshifts of all optical components near the transient are consistent, indicating that they all form part of the same host galaxy. The spectral parameters fitted for the galaxy core are listed in Appendix \ref{app:mw} Table \ref{Table:Optical_Spectra_fits}. 
The galaxy exhibits prominent emission lines, including [\ion{O}{2}], [\ion{O}{3}], H$\beta$, and H$\alpha$ (see Fig. \ref{fig:combined_overlay_spectra} and Appendix \ref{app:mw}), consistent with active star formation and relatively low dust extinction along the line of sight \citep{1998ApJ...498..541K,2000ApJ...533..682C}. The presence of high-excitation lines such as [\ion{Ne}{3}], as well as hydrogen and helium recombination features, suggests intense ionising radiation, potentially from vigorous star-forming activity or a low-luminosity AGN \citep{2002ApJS..142...35K,1981PASP...93....5B}. In star-forming environments, these emission lines trace ionised gas in \htwo\ regions and serve as important diagnostics of the physical conditions, including temperature, density, and chemical abundance patterns \citep{osterbrock_astrophysics_2006}.

We derived key physical parameters using a combination of broadband photometry and optical spectroscopy (see Appendix~\ref{Appendix:Gal_calcs} for details). These parameters are summarised in Table~\ref{tab:galaxy_params}. Based on the total stellar mass of \(1.55 \times 10^9~M_\odot\), the galaxy falls within the low-mass galaxy regime ($\lesssim10^9 M_\odot$) and is consistent with typical star-forming, disk-dominated systems \citep{2003MNRAS.341...54K}. The relatively low mass, combined with the derived star formation rate (\( 4.16~M_\odot\,\mathrm{yr}^{-1}\)), indicates an active star-forming galaxy. Star formation rates derived using the Calzetti attenuation law from extinction-corrected H$\alpha$ and UV luminosities are lower, at 0.40 and 2.24~$M_\odot\,\mathrm{yr}^{-1}$, respectively.\footnote{The discrepancy between these values and the higher SED-based star formation rate is expected, as different indicators probe star formation over distinct timescales and are subject to varying uncertainties in dust attenuation corrections \citep[see][]{2012ARA&A..50..531K}). SED fitting provides a time-averaged estimate of the star formation rate \citep{2013ARA&A..51..393C}.
} We find a Balmer decrement of 4.00, significantly higher than the intrinsic value of 2.86 \citep{1989agna.book.....O}, suggesting the presence of dust extinction likely associated with recent or ongoing star formation.

\begin{deluxetable}{lc}
\tablecaption{Physical Parameters of the Host Galaxy Core \label{tab:galaxy_params}}
\tablehead{
\colhead{Quantity} & 
\colhead{Value} 
}
\startdata
Stellar Mass ($M_\odot$)           & $1.55^{+0.42}_{-0.30} \times 10^9$ \\
SFR ($M_\odot\,\text{yr}^{-1}$)    
& $4.16^{+4.95}_{-1.52}$ \\
sSFR (Gyr$^{-1}$)                
& $0.39^{+0.34}_{-0.17}$ \\
Metallicity ($Z_\odot$) &  $0.58^{+0.42}_{-0.24}$ \\
Extinction Magnitude               & 
$1.16$ \\
Balmer Decrement                   & 4.00 \\
Ionisation Ratio                   & 
0.92 \\
\enddata

\tablecomments{Physical parameters were derived using extinction laws for star-forming \citep{2000ApJ...533..682C} galaxies. Core properties include stellar mass, gas mass, total mass, star formation rate (SFR), specific SFR (sSFR), metallicity, extinction magnitude, Balmer decrement and ionisation ratio.}
\vspace{-1cm}
\end{deluxetable}

The dust extinction corrected [O\,\textsc{iii}]/[O\,\textsc{ii}] ratio (\(\lesssim 1\)) points to a moderately low ionization environment, characteristic of evolved H\,\textsc{ii} regions rather than the hard radiation fields typical of AGN \citep{2002ApJS..142...35K,2006MNRAS.372..961K,2024ApJ...962..104F}. Furthermore, key Baldwin–Phillips–Terlevich \citep[BPT;][]{1981PASP...93....5B} diagnostic ratios are [\ion{N}{2}]/H$\alpha \lesssim 0.03$ (using $3\sigma$ limit on \ion{N}{2})  and [\ion{O}{3}]/H$\beta= 3.46$, consistent with the galaxy lying well within the star-forming region of the BPT diagram, characteristic of low-metallicity \ion{H}{2} regions ionized by young, massive stars, with no indication of AGN or composite excitation.
At the position of the transient, the KCWI spectrum reveals emission features, including [\ion{O}{2}] (3726\,\AA), $H\beta$ and [\ion{Ne}{3}] and the KCRM spectrum reveals emission features, including $H\alpha$ and [\ion{He}{1}] (7065\,\AA) (see Appendix \ref{app:mw} Fig. \ref{fig:optical-fits_transient_pos}). At the transient position, the Balmer decrement ($H\alpha/H\beta)$ is 4.34, significantly higher than the intrinsic value of 2.86 as well as the galaxy core, indicating a dusty region likely associated with star-formation.
Although the key Baldwin–Phillips–Terlevich \citep[BPT;][]{1981PASP...93....5B} diagnostic ratios—[\ion{N}{2}]/H$\alpha$ and [\ion{O}{3}]/H$\beta$—cannot be measured directly, upper limits derived from the continuum noise indicate that the source does not lie within the AGN regime. Furthermore, all detected emission lines have Gaussian dispersions of $\sigma < 1~\text{\AA}$, corresponding to narrow intrinsic velocities ($\sigma_v = c\,\sigma_\lambda / \lambda$; $\mathrm{FWHM} = 2.355\,\sigma_v$) of $\mathrm{FWHM} \lesssim 100~\mathrm{km~s^{-1}}$ for H$\alpha$,  H$\beta$, and \ion{O}{2}~$\lambda3726$. These low widths are consistent with photoionized gas in H\,\textsc{ii} regions or nuclear star clusters and are far below the several-hundred to thousand~km~s$^{-1}$ lines typical of AGN.

\begin{figure*}[t]
\gridline{\fig{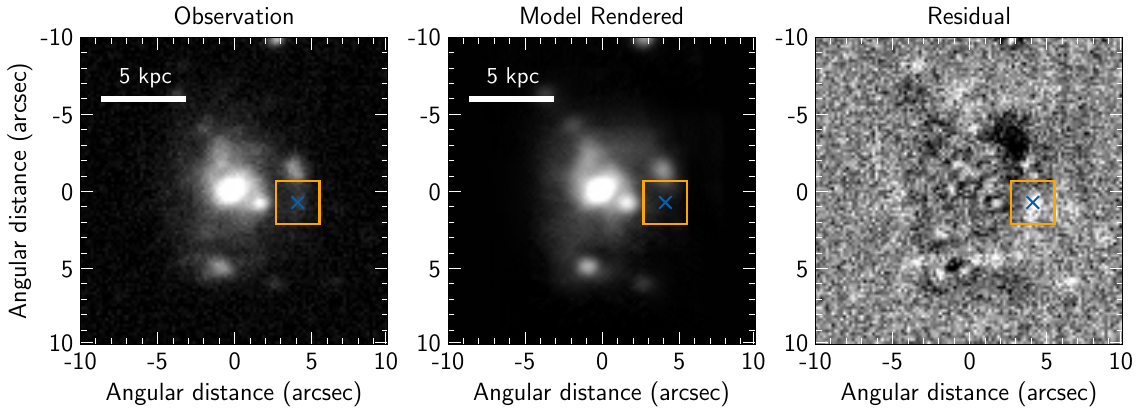}{1\textwidth}{Without model for NSC}} 
 \gridline{\fig{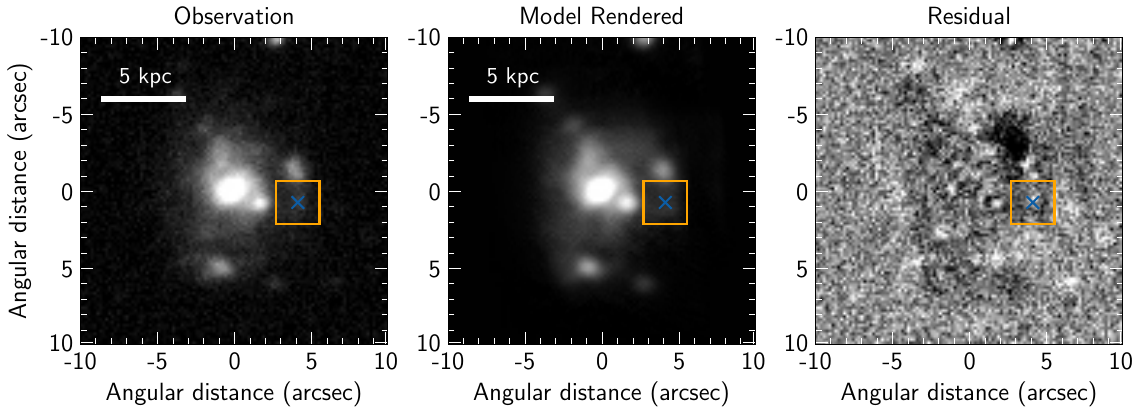}{1\textwidth}{With model for NSC}}
\caption{\texttt{Scarlet} scene model from Magellan imaging. Each source was modelled as a monotonically decreasing profile. We show the Magellan image (left), model rendered to match the Magellan imaging (centre), and the residual (right). Above, we show the model, observation and residuals for the case where no source is modelled at the position of the radio transient. Below, we show the case where a source is modelled at the radio transient position. We show the ATCA position centre with a blue cross, and an orange box where an additional extended source was fit to account for galaxy emission in the region of the transient.}
\label{fig:Scarletmodel}
\end{figure*}

Taken together, the spatial coincidence, low chance alignment probability, spectroscopic diagnostics, and derived physical parameters strongly support the interpretation that 2dFGRS TGS143Z140 is the likely host galaxy of \source\ and strongly disfavour a background AGN as a plausible explanation. 

\subsection{Imaging}

\subsubsection{IR}
\label{ss_IR_hg}
No infrared counterpart was detected in the stacked image above the 5-sigma infrared detection threshold of m(J) = 23.3 in AB magnitudes. The stacked image is presented in Fig. \ref{fig:combined_overlay_spectra}.


\subsubsection{Optical}

The deep Magellan imaging of the galaxy reveals a complex post-merger system with multiple extended components. Keck spectroscopy confirms that all of these structures lie at the same redshift, with the outer clumps consistent with \ion{H}{2} regions. The galaxy exhibits prominent tidal tails, indicative of a recent interaction or merger. This interpretation is further supported by spatially resolved velocity maps, which show a smooth and continuous emission-line velocity gradient along the tidal features (see Fig. \ref{fig:velmap} in Appendix \ref{app:mw}). No kinematically distinct component is detected at the location of the transient, indicating that the gas in this region is dynamically connected to the host galaxy and its tidal debris rather than associated with a background system. The radio transient is located at the end of a tidal tail, coincident with a slight enhancement in emission.

To assess emission excess at the transient location, we modelled the Magellan data using the \texttt{Scarlet} scene modelling software\footnote{\url{https://pmelchior.github.io/scarlet/}} \citep[][]{Melchior2018}. We constructed the PSF model by weighted-mean stacking cutouts of all $>10\sigma$ stars in the image. On a $20^{\prime\prime}$ by $20^{\prime\prime}$ cutout centred on the host galaxy, we used Source Extractor \citep{Bertin1996} to detect all sources above $2.5\sigma$. We modelled each source as a non-parametric, monotonically decreasing (but not radially symmetric) extended source and ran \texttt{Scarlet} to convergence. The best-fit model, together with the observations and residuals, is shown in Fig. \ref{fig:Scarletmodel}. A clear flux excess relative to the \texttt{Scarlet} model of the host galaxy is evident at the position of the radio transient. To place an upper limit on the magnitude of this excess, we refit the model including an additional extended source component at the transient location (Fig.~\ref{fig:Scarletmodel}). The best-fit AB magnitude and $5\sigma$ uncertainty of the excess emission is $m_g = 24.1 \pm 0.7$. This enabled us to estimate the flux of local galaxy emission in the region of the transient. We also used the pixel variance of the residuals in a 30 × 30 pixel cutout at the source location to determine that the $5\sigma$ limiting magnitude based on the background variance was 28.0 (AB magnitudes).

\section{Transient Classification and Modelling}
\label{sec:classification}

In this section, we estimate the physical parameters of \source\ and compare with other transient classes to help its classification.

\subsection{Spectral Luminosity and Lightcurve Evolution}

\begin{figure*}[t]
    \centering
    \includegraphics[width=0.8\linewidth]{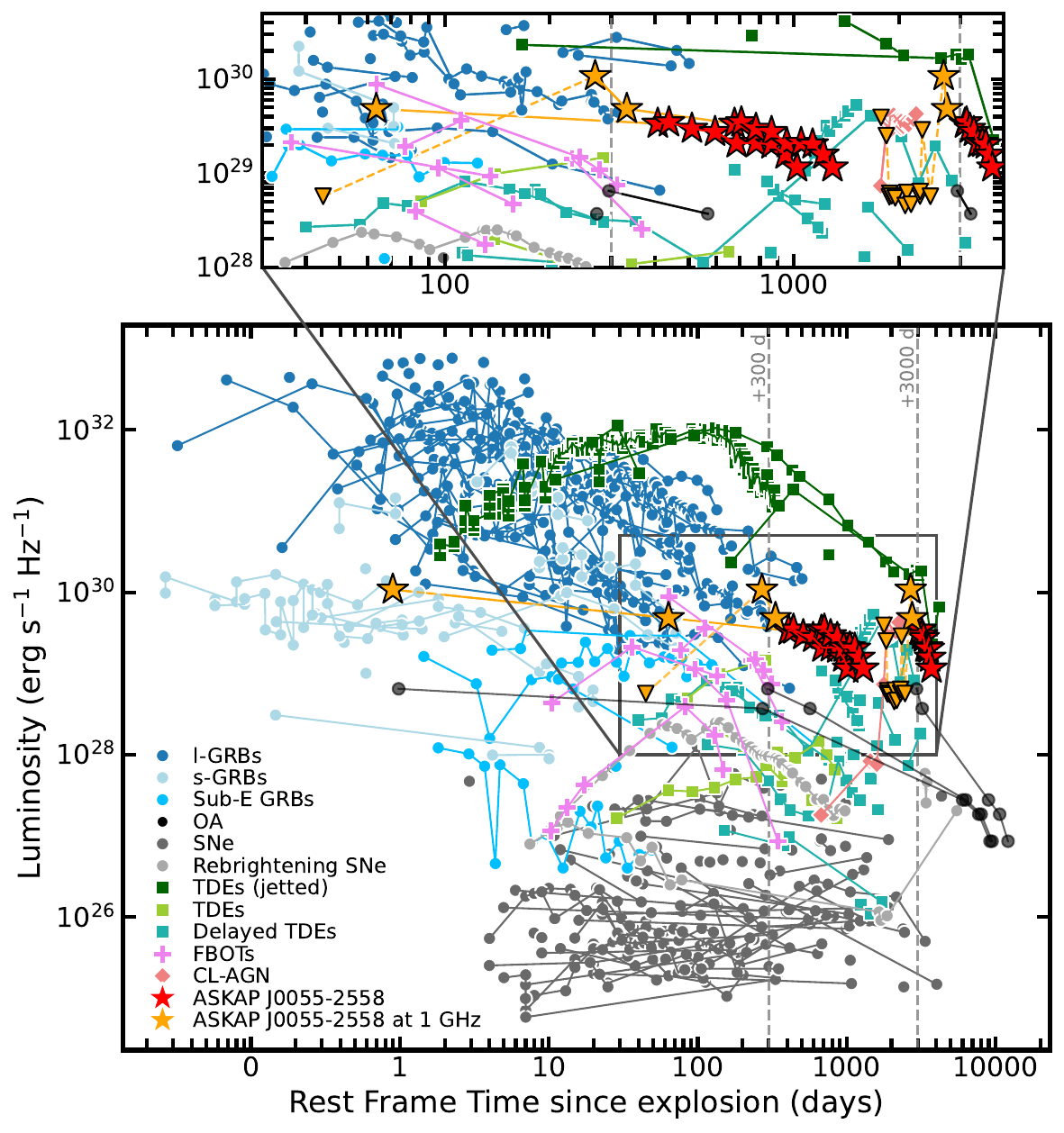}
   
    \caption{8–10 GHz light curve of \source\ (red stars) shown in comparison with other classes of explosive transients: gamma-ray bursts (long-GRBs; blue circles), short-GRBs (light-blue circles), sub-energetic GRBs (deep sky blue circles), supernovae (SNe; dark gray circles), rebrightening SNe (light gray circles), tidal disruption events (TDEs; light-green squares), relativistic TDEs (dark-green squares), delayed TDEs (light seagreen squares), fast blue optical transients (FBOTs; pink pluses), and changing-look AGN (coral diamonds). Luminosities are plotted relative to 2022 Feb 3 (one day before first detection), though the true explosion date may precede this; we therefore show \source\ with additional shifted reference times of 300 and 3000 days prior to detection. Orange datapoints represent the early-time low frequency radio emission of \source\ at 1\,GHz. 5-sigma limits are shown using downward triangles. No limits are shown for observations assuming launch was a day before first radio detection. Jetted TDE AT2022cmc's datapoints are plotted at 15.5\, GHz \citep{2025arXiv250613618R}. The source's luminosity is comparable to late-time GRBs and TDEs, and exceeds that of SNe and thermal TDEs. For context, another orphan radio afterglow candidate is shown in black circles, time-shifted similarly to \source. To recreate this plot, please use code \url{https://github.com/ashnagulati/Transient_Comparison_Plots.git} \dataset[]{https://doi.org/10.5281/zenodo.17774690}. References: \citet{2012ApJ...746..156C,alexander_discovery_2016,alexander_radio_2017,andreoni_very_2022,berger_grb_2001,berger_host_2001,berger_jet_2000,berger_radio_2003,bietenholz_radio_2021,bright_radio_2022,brown_late-time_2017,cendes_mildly_2022,cendes_radio_2021,cendes_ubiquitous_2024,cenko_afterglow_2011,cenko_multiwavelength_2006,cenko_swift_2012,chandra_comprehensive_2008,chandra_discovery_2010,chrimes_multi-wavelength_2024,coppejans_mildly_2020,djorgovski_afterglow_2001,eftekhari_associating_2018,frail_accurate_2005,frail_energetic_2006,frail_enigmatic_2000,frail_radio_1999,galama_bright_2000,galama_continued_2003,goodwin_radio_2023,goodwin_systematic_2025,greiner_unusual_2013,harrison_broadband_2001,harrison_optical_1999,ho_at2018cow_2019,ho_koala_2020,horesh_are_2021,horesh_delayed_2021,horesh_unusual_2015,laskar_radio_2023,laskar_reverse_2016,laskar_vla_2018,2014ApJ...780..118F,2022ApJ...935L..11L,2019ApJ...883...48L,2021ApJ...906..127F,2024ApJ...975L..13A,2006ApJ...650..261S,2024Natur.626..737L,2005Natur.438..988B,2021ApJ...906..127F,anderson2025,law_discovery_2018,leung_search_2021,margutti_embedded_2019,margutti_signature_2013,mattila_dust-enshrouded_2018,meyer_late-time_2025,moin_radio_2013,mooley_late-time_2022,oconnor_structured_2023,pasham_multiwavelength_2015,perley_afterglow_2014,perley_grb_2008,rhodes_rocking_2024,rol_grb_2007,rose_late-time_2024,schroeder_long-lived_2025,schroeder_radio_2024,sfaradi_off-axis_2024,soderberg_constraints_2004,soderberg_redshift_2004,soderberg_relativistic_2006,stein_tidal_2021,taylor_discovery_1998,van_der_horst_detailed_2008,zauderer_radio_2013,2017ApJ...837..153A,2024A&A...691A.329C,1998Natur.395..663K,1997Natur.389..261F,2025ApJ...983...29H,2012GCN.12804....1H,2023GCN.33475....1A,2008ApJ...688..470P,2002ApJ...572L..51P,2008A&A...480...35V}}
     \label{Figure:Lum}
\end{figure*}

In Fig.~\ref{Figure:Lum}, we compare the luminosity evolution of \source\ for three possible start dates with that of other synchrotron transients. If \source\ is indeed associated with the galaxy 2dFGRS TGS143Z140 ($z = 0.116$), its high luminosity is unprecedented among known supernovae \citep{2021ApJ...908...75B} and thermal TDEs\footnote{We define ``thermal'' TDEs as events that exhibit an early, single radio flare associated with the initial accretion phase, and ``delayed'' TDEs as those in which the radio emission either shows a second, later flare or appears only months to years after disruption.} \citep{2022MNRAS.511.5328G,2023MNRAS.518..847G}. Additionally, the transient exhibits a slow decline between $\sim$400-1000 days post first radio-detection. This long-lived, gradual evolution contrasts with the behaviour of FBOTs, which typically fade much more rapidly, often at rates of $\gtrsim t^{-3}$ and complete their entire evolution in $<1000$\, days \citep{2019ApJ...872...18M,2020ApJ...895L..23C,2020ApJ...895...49H}.

Its luminosity evolution appears to resemble the late-time emission from a jetted TDE like SWIFT J164449.3+573451 \citep{2018ApJ...854...86E}, or a long-GRB (l-GRB). Alternatively, luminosity may also be consistent with delayed TDEs, where flares occur long after the initial optical burst as seen in AT2018hyz \citep{2022ApJ...938...28C,sfaradi_off-axis_2024}. In terms of longevity, \source\ is consistent with off-axis GRB afterglows, such as that of GW170817 \citep{2025MNRAS.539.2654K}, and with TDE events that have entered their late-time Newtonian phase, such as AT2019azh \citep{2025ApJ...993..207B} and AT 2022cmc \citep{2025arXiv250613618R}. However, the lack of early-time coverage complicates efforts to determine whether this was an on-axis or off-axis event. At late times, distinguishing between the two is challenging, as both exhibit similar evolution once the jet decelerates, widens, and becomes Newtonian \citep{2002MNRAS.332..735H,2018MNRAS.481.2581L}.

Taken together, the luminosity evolution is inconsistent with supernovae, thermal TDEs, FBOTs, and even lower-luminosity sources such as X-ray binaries \citep{1983ApJ...273L..65G,1996ASPC...93..252S} and radio bubbles \citep{2020ApJ...896..117B,2019MNRAS.482.2389U}. Instead, it aligns more closely with the late-time behaviour of a l-GRB or a jetted/delayed TDE.

 \subsection{Equipartition Analysis}
\label{ss_eqipartition_analysis}

\begin{figure*}[t]
    \centering
    \includegraphics[width=0.90\textwidth]{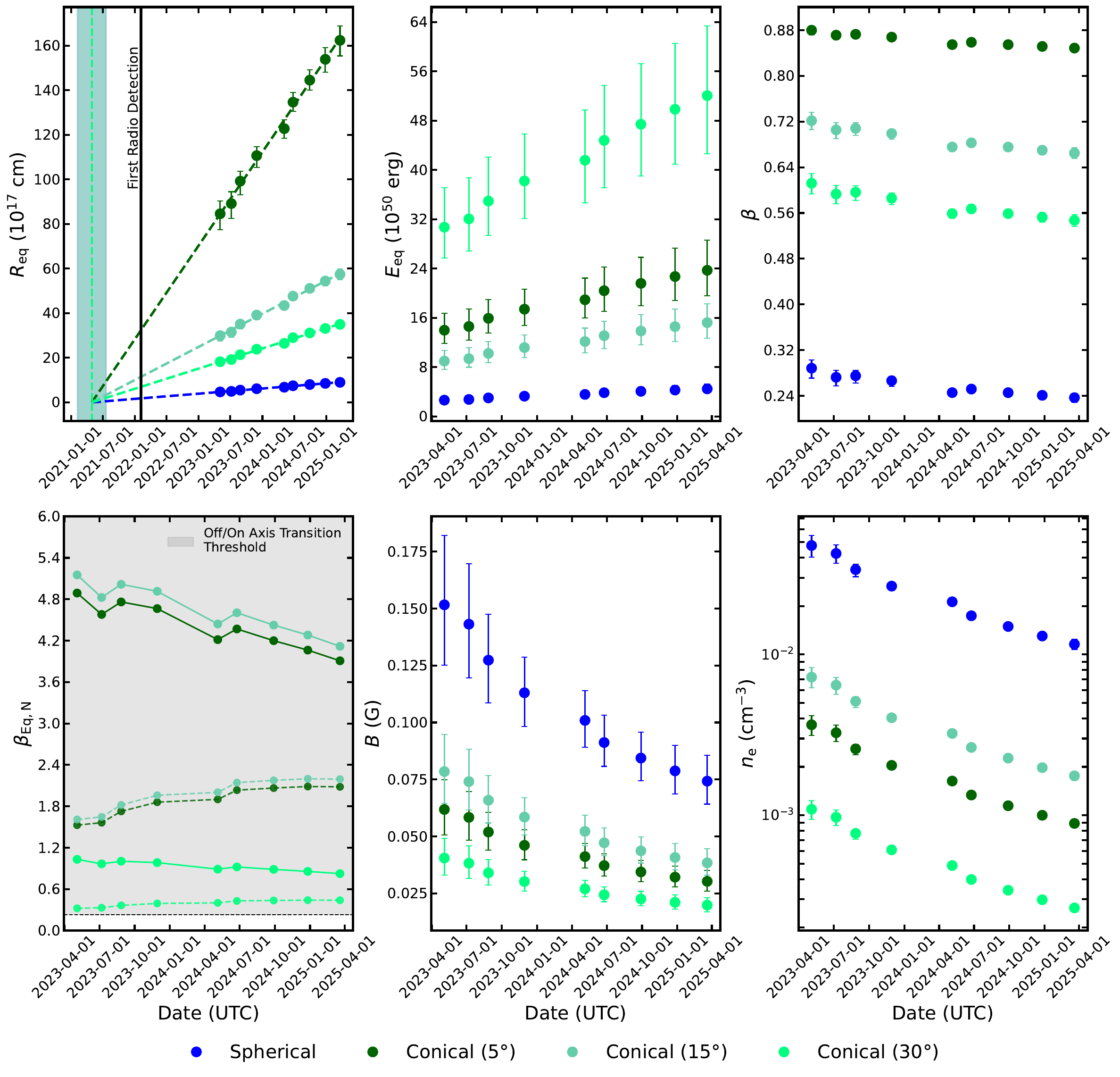}
   
    \caption{Outflow properties of ASKAP J0055–2558 inferred from equipartition modelling of its radio spectra from epoch 2023-08-27 to 2025-03-20. Results assuming spherical geometry are shown in blue, and conical geometry in shades of green. Shown are the outflow radius ($R$), energy ($E$), velocity ($\beta$), equipartition velocity ($\beta_{eq,N}$), magnetic field strength ($B$) and ambient electron density ($n_e$). E and R are derived under equipartition and are uncorrected for deviations from it.  In the upper left panel, dashed lines show linear fits to R for each geometry. The first radio detection is marked by a solid black line; dashed vertical lines (colour-coded by geometry) indicate the estimated outflow launch date (when $R = 0$). In the lower left panel, Equipartition-based Newtonian velocity ($\beta_{\rm eq,N}$) versus epoch of observation. Since $\beta_{\rm eq,N} \propto t^{-1}$, dashed lines show the evolution assuming an outflow launch 1000 days prior to detection. The grey shaded region marks the on/off-axis threshold; values above 0.23 suggest the outflow may have originated as a relativistic jet that has since decelerated into the Newtonian regime.}
    \label{Figure:Outflow_params}
\end{figure*}

To constrain the physical parameters of the synchrotron-emitting region, we adopt the synchrotron self-absorption model developed by \citet{2013ApJ...772...78B}. This model, which assumes an instantaneous energy injection event, relates observed spectral properties (peak flux density $F_{p}$ and peak frequency $\nu_p$) to physical parameters based on the assumption of equipartition between the energy densities of relativistic electrons and magnetic fields. The parameters minimal energy radius ($R_{\text{eq}}$) and the total minimal energy ($E_{\text{eq}}$) are calculated using the relativistic solutions presented in Equations 27 and 28 from \citet{2013ApJ...772...78B}. Crucially, in the absence of a measured $\nu_m$ (the synchrotron frequency of the minimal Lorentz factor electrons), we assume $\eta = 1$, corresponding to the simplified case where the synchrotron self-absorption frequency $\nu_a$ is greater than or equal to $\nu_m$ ($\nu_a \geq \nu_m$).

Since the bulk Lorentz Factor ($\Gamma$) of the outflow is not directly observed, we adopt the theoretical value derived from the shock acceleration model outlined in \citet{2013ApJ...772...78B}. The model relates the minimal electron Lorentz factor, $\gamma_m$, to the source Lorentz Factor $\Gamma$ via $\gamma_m = \chi_e(\Gamma - 1)$, where $\chi_e$ depends on the electron power-law index $p$ and the fraction of the shock energy imparted to the electrons, $\epsilon_e$:
$$
\chi_e = \left(\frac{p-2}{p-1}\right) \epsilon_e \left(\frac{m_p}{m_e}\right)
$$
By assuming a fixed value for the electron energy fraction, $\epsilon_e = 0.1$, and setting $\gamma_m = 2$, we derive a fixed, theoretical value for the bulk Lorentz factor $\Gamma$ for the source, which is then used as a parameter in the calculation of $R_{\text{eq}}$ and $E_{\text{eq}}$ via the \citet{2013ApJ...772...78B} equations.

The geometric factors, $f_A$ (area factor) and $f_V$ (volume factor), are applied based on the assumed geometry. For spherical geometry, $f_A = 1.0$ and $f_V=4/3$ \citep{2013ApJ...772...78B}. For conical geometry, the area factor $f_A$ is set based on the half-opening angle $\theta_j$. For narrow angles ($\theta_j \leq 10^\circ$), we implement the relativistic jet convention where $f_A = f_V = \Gamma \theta_j^2$. For wider angles, we use the geometric expression $f_A=1-\cos\theta_j$ and fix the volume filling factor at $f_V=1.0$, following the standard unit-volume convention in non-relativistic equipartition models \citep{2016ApJ...819L..25A,2024MNRAS.528.7123G}. For opening angles($\Theta$) of $5^\circ$, $15^\circ$, and $30^\circ$, this yields $f_A=f_V=0.0019$, $f_A=f_V=0.017$, and $f_A=0.034$ with $f_V=1.0$, respectively. 

Following the calculation of $R_{\text{eq}}$ and $E_{\text{eq}}$, the magnetic Field ($B$), electron number ($N_e$) and ambient density ($n_e$) are calculated directly from the equipartition relations. The outflow velocity ($\beta_{\text{ej}}$) is then determined kinetically. This velocity represents the actual speed of the blastwave's shock front at the time of observation $t$. In this case, $t$ has been calculated relative to the first radio detection. This $\beta_{\text{ej}}$ is calculated by solving the expression that relates the instantaneous radius $R_{\text{eq}}$ and the observed time:
$$
\frac{R_{\text{eq}}(1+z)}{c t} = \frac{\beta_{\text{ej}}}{1 - \beta_{\text{ej}}}
$$
The parameters are summarised in Fig. \ref{Figure:Outflow_params} and in Appendix \ref{app:mw} Table \ref{Table:equi_params}. Since the burst (or explosion) date is unknown but must precede the first radio detection, it should be noted that only the velocity ($\beta \propto t^{-1}$) varies explicitly with time, and we have illustrated how this variation would appear. We also calculated the relative velocity for each geometry and epoch, and we find that while there are large errors on these numbers, they are consistent with the velocity values extrapolated using the time of first radio detection as the launch date. Furthermore, astrophysical sources may deviate from equipartition, hence, we applied corrections assuming plausible ranges for electron, $\varepsilon_e \in \{0.1, 0.0005\}$, and magnetic energy fractions, $\varepsilon_B \in \{0.1, 0.01, 0.001\}$ \citep{2013MNRAS.436.1258H,2016ApJ...819L..25A,2020ApJ...896..166R,2021ApJ...908..125C,2022MNRAS.511.5328G} and the following equations.

\vspace{-0.3cm}

\begin{align}
\epsilon &= \left( \frac{\epsilon_B}{\epsilon_e} \right) \left( \frac{11}{6} \right) \\[8pt]
R &= R_{\rm eq} \cdot \epsilon^{1/17} \\[8pt]
E &= E_{\rm eq} \cdot \left[ \frac{11}{17} \cdot \epsilon^{-6/17} + \frac{6}{17} \cdot \epsilon^{11/17} \right]
\end{align}

In the following sections, we compare these parameters to other transient classes.

\subsubsection{Energy-velocity}

\begin{figure*}[htbp]
    \centering
    \includegraphics[width=\linewidth]{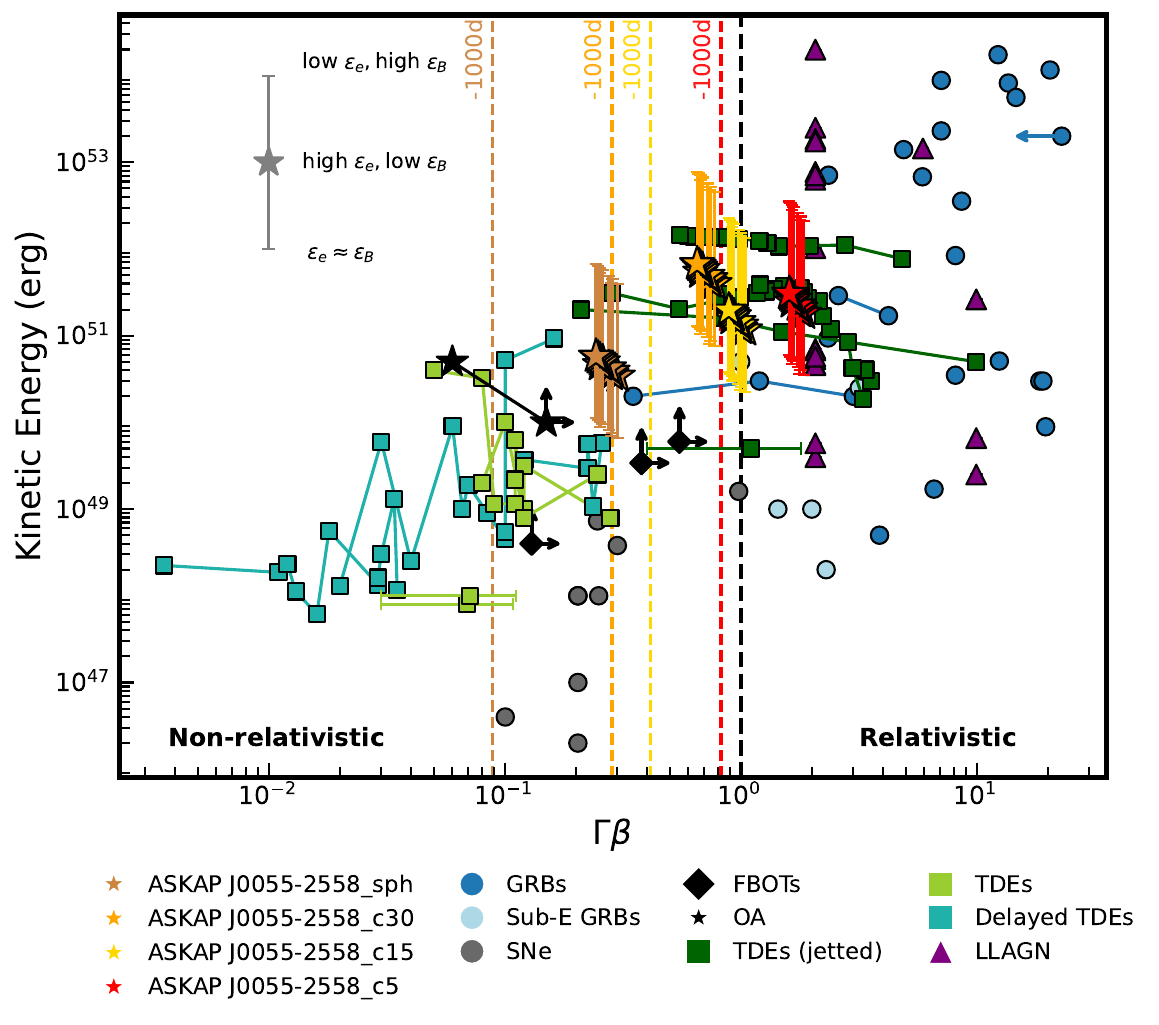}
   
    \caption{Intrinsic (beaming-corrected) kinetic energy ($E_{\rm K}$) versus outflow speed ($\Gamma\beta$), where $\Gamma = (1 - \beta^2)^{-1/2}$ is the bulk Lorentz factor and $\beta = v/c$ is the velocity, computed with time referenced to the first radio detection on 2022 Feb 4. \source\ is shown assuming different conical geometries: red stars (5°), yellow (15°), orange (30°), and brown (spherical). Stars (from right to left) mark nine observer-frame epochs. For comparison, other explosive transients are included: GRBs (blue circles), sub-energetic GRBs (light-blue), SNe (dark grey), TDEs (light-green squares), relativistic TDEs (dark-green), delayed TDEs (light sea-green), FBOTs (black diamonds), LLAGN (purple triangles), and a radio orphan afterglow candidate (black stars). All values are derived from radio observations; for LLAGN, constant jet power over 10 years is assumed. The error bars on energy values for \source\ reflect the range of corrections for deviations from equipartition, based on combinations of $\varepsilon_e$ and $\varepsilon_B$. As the explosion time is uncertain and $\beta \propto t^{-1}$, velocities are scaled assuming an explosion 1000 days before first detection; dashed lines indicate this assumed shift.  To recreate this plot, please use code \url{https://github.com/ashnagulati/Transient_Comparison_Plots.git}. References: \citet{2008Natur.453..469S,2010ApJ...725..922S,2005ApJ...621..908S,2013arXiv1309.3573K,2015ApJ...805..187C,2006ApJ...651.1005S,2002ApJ...577L...5B,2013MNRAS.428.1207S,2005ApJ...621..908S,2012ApJ...752...78S,2019ApJ...870...38S,2006Natur.442.1014S,2013ApJ...778...18M,1998Natur.395..663K,2003Natur.426..154B,2001ApJ...554..667P,2003ApJ...597..459Y,2000ApJ...545...56B,2006ApJ...650..261S,2005Natur.437..845F,2005Natur.438..988B,2010ApJ...711..641C,2006ApJ...646L..99F,2008ApJ...683..924C,2011ApJ...732...29C,2013ApJ...776..119L,2014ApJ...781....1L,2005ApJ...619..994F,2008A&A...480...35V,2018ApJ...854...86E,2018Sci...361..482M,2012ApJ...753...77C,2016ApJ...819L..25A,2023MNRAS.518..847G,2022MNRAS.511.5328G,2017ApJ...837..153A,2022ApJ...938...28C,2025ApJ...981..122G,2025ApJ...983...29H,2019ApJ...871...73H,2020ApJ...895L..23C,2020ApJ...895...49H,2014ApJ...780L..14M, 2022ApJ...924...16M,2025arXiv250613618R}}
    \label{Figure:E_gammabeta}
\end{figure*}

The inferred outflow velocity from the equipartition analysis in Section \ref{ss_eqipartition_analysis}, $\beta$, spans $0.2-0.9$, with higher values for conical geometries, placing \source\ in the mildly relativistic regime. These estimates, anchored to the first radio detection, represent upper limits, as the outflow launch likely preceded the first radio detection. The corresponding equipartition-corrected energy ranges from $10^{50}$ to $10^{53}$~erg. We find that the inferred outflow velocity shows clear mild deceleration in the spherical model, while the conical models are consistent with only weak deceleration and nearly constant expansion speed. The observed decreasing velocity follows expectations for jetted TDEs (e.g., Swift~J1644+57; \citealt{2018ApJ...854...86E}) and GRBs (e.g., GRB 030329; \citealt{2008HEAD...10.0506V}). In all geometries, the equipartition energy increases monotonically with time, indicative of ongoing energy injection such as slower ejecta catching up with the forward shock, which appears to taper off and approach a plateau at the latest epochs. This flattening occurs when the outflow has swept up an external mass comparable to its own and begins to decelerate, causing the radio emission to fade as the kinetic energy stabilises \citep{2015MNRAS.448..417B}.

In Fig.~\ref{Figure:E_gammabeta}, we place \source\ in energy-velocity space alongside various synchrotron transient classes. Its kinetic energy and velocity significantly exceed those of ordinary core-collapse supernovae ($E_k \sim 10^{51}$~erg; $\beta \ll 1$), but remain below those of typical, on-axis GRBs. It lies within the mildly relativistic, high-energy regime characteristic of the late-time evolution of jetted TDEs and l-GRBs, delayed TDEs and FBOTs. The mildly relativistic velocities and high kinetic energy suggest \source\ to be the late-time phase of an engine-driven transient, such as the one found by \citet{2018ApJ...866L..22L}, claimed to be an off-axis orphan afterglow event. 

The distribution of kinetic energy as a function of outflow velocity ($E_{eq}$--$\Gamma\beta^k$) exhibits a steeply declining slope of roughly -2.5 to -3.7, indicating that the fastest-moving material carries comparatively little energy, found also in the Newtonian afterglow of GW170817 \citep{2022ApJ...938...12B,2025MNRAS.539.2654K} and on-axis jetted TDE SWIFT J1644+57 \citep{2018ApJ...854...86E}. However, inferred velocity depends inversely on the assumed launch time. Adopting a delay of 300–1000  days prior to first radio detection shifts the outflow towards a comparatively less relativistic regime ($\Gamma\beta < 1 $; indicated by the dashed lines in Fig. \ref{Figure:E_gammabeta}), where the relationship transitions towards a positive slope with increasing delay, consistent with the Sedov–Taylor scaling $E \propto v^{2}$. This supports a scenario where the outflow has significantly decelerated and is in the deep Newtonian regime. Either of those scenarios would still place \source\ in the non-relativistic phase of a TDE or GRB.



\subsection{Host Galaxy and Local Environment}
\label{ss_host_disc}

The host galaxy of this transient is a low-mass, moderately star-forming, and dusty system, characterised by prominent \ion{H}{2} regions. Galaxies with similar properties are known to host a range of energetic transients, including supernovae, GRBs, and FBOTs \citep{2020ApJ...895L..37B,2020ApJ...895L..23C}. The transient lies at a projected distance of $\sim$8.7 kpc from the galaxy’s core, larger than typical offsets for long GRBs, but it is coincident with a star-forming region in the galaxy’s outskirts. Given that long GRBs are preferentially associated with sites of active star formation, and that the host shows a complex morphology with extended \ion{H}{2} regions, an orphan GRB afterglow remains a plausible explanation despite the relatively large offset.

Deep Magellan imaging also reveals an additional flux excess spatially coincident with the transient position within the positional uncertainty. While this excess lies above the radio-to-optical extrapolation of the optically thin synchrotron spectrum, it could be associated with the transient if the optical emission is dominated by thermal radiation from accretion-powered reprocessing or outflows \citep{2016MNRAS.461..948M,2016ApJ...827....3R}, rather than by a jet afterglow. Alternatively, the excess emission at the transient position may originate from a globular or nuclear star cluster with an inferred absolute magnitude of $M_g \approx -14.71$ (uncorrected for extinction), potentially hosting a massive black hole (MBH). This estimate should be regarded as an upper limit, as the detected emission could alternatively arise from a clumpy tidal feature associated with the primary host galaxy.

Using the measured flux of this excess and assuming it arises from a star cluster, we estimate the stellar mass under three different age hypotheses. From this measured $g$-band magnitude, we derived stellar masses by combining rest–frame $g$-band luminosity with mass-to-light ratios from \texttt{EzGal}\footnote{\url{https://github.com/cmancone/easyGalaxy}} SSP models (Chabrier IMF, solar metallicity), yielding $M_\star \sim 3\times10^{6}$, $4\times10^{7}$, and $3\times10^{8}\,M_\odot$ for ages of 10 Myr, 1 Gyr, and 10 Gyr, respectively (the 10 Myr value corresponds to the youngest valid SSP grid point). At the low-mass end ($M_\star \sim 3\times10^{6}\,M_\odot$, young), the system would resemble a ``super star cluster'' in a starburst galaxy such as M82 \citep{2007ApJ...663..844M,2010ARA&A..48..431P}. At the high-mass end ($M_\star \sim 10^{8}\,M_\odot$, old), it approaches the regime of the most massive ultra-compact dwarfs (UCDs; \citealt{2007A&A...463..119H,2011MNRAS.412.1627C,2008A&A...487..921M,2015ApJ...812L...2L}), while the overall mass range overlaps with that of nuclear star clusters \citep{2005ApJ...618..237W}. For context, extreme Virgo UCDs such as M59-UCD3, with a luminosity comparable to our candidate, and M60-UCD1 both host central black holes of a few $\times10^{6}\,M_\odot$ \citep{2014Natur.513..398S,2015ApJ...812L...2L,2018ApJ...858..102A}.

To estimate the central black-hole mass in this potential star cluster, we apply the star cluster scaling relation of \citet{2020MNRAS.492.3263G} to the young and intermediate-age cases, obtaining $M_{\rm bh}\sim4.8\times10^{4}\,M_\odot$ (10 Myr) and $\sim3.8\times10^{7}\,M_\odot$ (1 Gyr). For the old case ($M_{\rm nc}\sim3.4\times10^{8}\,M_\odot$), the relation would imply $M_{\rm bh}\sim1.2\times10^{10}\,M_\odot$, unphysically exceeding the cluster mass by a factor of $\sim30$, so we consider it inapplicable. The high luminosity and stellar mass disfavour a classical globular cluster and instead point to a nuclear star cluster, consistent with environments expected to host $10^{4}$–$10^{7}\,M_\odot$ black holes and supporting the interpretation of a TDE.


\subsection{Ambient Electron Density-Radius}

\begin{figure}[t]
    \centering
    \includegraphics[width=\linewidth]{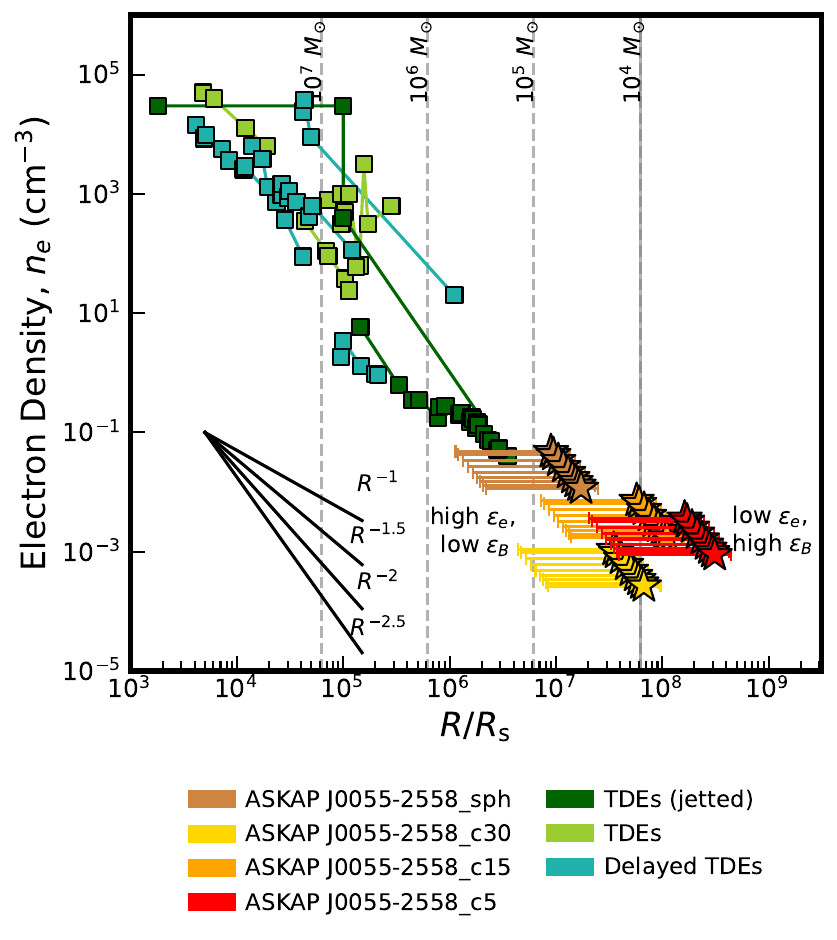}
   
    \caption{Variation of ambient density with distance from the black hole for TDEs compared to \source\. Distances are scaled to Schwarzschild radii. \source\ is shown assuming different conical geometries: red stars (5°), yellow (15°), orange (30°), and brown (spherical). The radii are corrected for deviations from equipartition, with uncertainties reflecting the range of $\varepsilon_e$ and $\varepsilon_B$ combinations. The grey dashed vertical lines show distance scalings for our source assuming black holes of masses between $10^4$ and $10^7\,M_\odot$ as derived in Section \ref{ss_host_disc}.
    References: \citet{stein_tidal_2021,cendes_radio_2021,alexander_discovery_2016,eftekhari_radio_2018,mattila_dust-enshrouded_2018,alexander_radio_2017,goodwin_at2019azh_2022,goodwin_radio_2023,goodwin_second_2024,2025ApJ...983...29H,2022ApJ...938...28C}}
    \label{Figure:R_ne}
\end{figure}

In Fig. \ref{Figure:R_ne}, we place ASKAPJ0055$-$2558 in electron-density–radius space alongside known TDEs. For this comparison, we adopt black hole masses of $10^{4}$–$10^{7}\,M_\odot$ (Section~\ref{ss_host_disc}), with the intermediate-mass range showing the best consistency with the observed trend. The inferred $\log(n_e)$–$\log(R_{\rm eq})$ trends across all assumed geometries indicate a decreasing density profile, with power-law slopes of $\sim-2.2$. This trend is consistent with that observed in the TDE population (Fig. \ref{Figure:R_ne}), in core-collapse supernovae \citep[e.g.,][]{1982ApJ...259..302C, 2012ApJ...755..110C}, and in GRBs expanding into a stellar-wind environment \citep{2000ApJ...543...66P, 2000ApJ...536..195C}. Similar behaviour is seen in GRB 171205A, where the afterglow remains consistent with a wind-like medium even beyond 1000 days post-burst, showing no evidence for a transition to a constant-density regime \citep{leung_search_2021, 2021ApJ...907...60M}.

Notably, the circumnuclear medium (CNM) surrounding \source\ appears significantly less dense than that typically associated with thermal TDEs, FBOTs \citep[e.g.,][]{2019ApJ...872...18M, 2020ApJ...895L..23C}, and supernovae \citep{1996ApJ...461..993F, 2006ApJ...651.1005S, 2012ApJ...755..110C}. The relatively low density inferred for \source\ is more consistent with GRB environments \citep{2002ApJ...571..779P, 2013ApJ...776..119L} or with the sparse surroundings which may be expected near an IMBH in a globular cluster \citep[for eg.,][]{2001ApJ...557L.105F}. This low density could explain the transient’s slow evolution. As shown by \citet{2020MNRAS.492..686L}, transients in denser environments tend to evolve faster and appear brighter due to stronger interaction between the outflow and the surrounding material. Overall, the observed density scaling across different geometries supports a scenario in which a jet is propagating into a low-density, steeply declining coherent circumburst medium.


\section{Orphan Afterglow or off-nuclear IMBH TDE: Physical Plausibility}
\label{sec:physical_plausibility}

In this section, we evaluate whether the transient classification inferred from our comparative analysis is physically plausible. Specifically, we assess whether the observed properties can be explained by synchrotron emission from shock-accelerated electrons.

GRBs are followed by long-lived, multi-wavelength afterglows produced as the relativistic blast wave decelerates in the surrounding medium, causing the Lorentz factor ($\Gamma$) to drop and the beaming angle to widen as $\Omega(t)\propto1/\Gamma(t)^2$ \citep{1997ApJ...482L..29M}. Even without a detected prompt GRB, the afterglow can become visible once the jet decelerates and spreads laterally, producing so-called orphan afterglows \citep{1999ApJ...513..656D}. These arise either because the GRB was outside the detector’s field of view or because the jet was initially misaligned with the observer ($\theta_{\rm view}>\theta_j$; \citealt{2003ApJ...591.1097R,2022Galax..10...93S}). While many optical \citep{2015ApJ...803L..24C,2020ApJ...905...98H} and a few radio candidates have been reported \citep{2023MNRAS.523.4029L}, only one convincing case has been identified in the radio \citep{2018ApJ...866L..22L,2019ApJ...876L..14M,2022ApJ...924...16M}.

TDEs occur when a star is disrupted by a black hole’s tidal forces \citep{1975Natur.254..295H,1988Natur.333..523R}, typically in galaxy nuclei hosting supermassive black holes (SMBHs; $M_\mathrm{BH}>10^6\,M_\odot$; \citealt{2020SSRv..216...81A}). Radio studies reveal two main classes: non-relativistic, thermal events \citep{2022MNRAS.511.5328G} and relativistic, non-thermal, jetted events \citep{2018ApJ...854...86E}. A rarer subclass involves IMBHs, seen in dwarf-galaxy nuclei \citep{2022NatAs...6.1452A} or in off-nuclear environments such as galaxy clusters \citep{2018NatAs...2..656L}. Although IMBH-associated TDEs have not yet been detected at radio wavelengths, several recent discoveries include radio afterglows from off-nuclear candidates \citep{2025NatAs...9..702Z,2025arXiv250109580J,2025arXiv250925877L,2025arXiv250922843E,2025ApJS..278...36G,2025arXiv250714286L,2025arXiv250803807S}, highlighting that off-nuclear events can indeed produce detectable radio emission.

Studies of orphan afterglow candidates in the last few decades, such as J060938-333508 \citep{bannister_22-yr_2011} and ASKAP J175036.1-181454 \citep{2023MNRAS.523.4029L}, demonstrate long-lived, luminous synchrotron emission with peaked or evolving spectra and off-nuclear host associations, key features also observed in ASKAP J0055-2558. These events, together with well-studied analogues such as the orphan afterglow candidate FIRST J141918.9+394036 \citep{law_discovery_2018} and the off-nuclear TDE AT~2024tvd \citep{2025arXiv250803807S}, highlight an emerging population of radio transients likely powered by relativistic outflows occurring at significant offsets from their host galaxy centres. The persistence of their radio emission over years, combined with the absence of contemporaneous high-energy counterparts, collectively suggests that ASKAP J0055-2558 is part of an emerging class of extragalactic, synchrotron-powered transients such as orphan afterglows or off-nuclear TDEs.

\subsection{Spectral and Temporal Index Consistency: Closure Relations}

We compare the derived spectral and temporal indices for \source{} with the standard closure relations for synchrotron afterglows in the Newtonian (wide-angle geometries) and relativistic (narrow-angle geometries) phase \citep[][]{2013NewAR..57..141G,2002ApJ...568..820G}. Based on the electron density–radius scaling, a wind-like environment is favoured. In such a medium, with spectral ordering $\nu_m < \nu_a < \nu_c$ and assuming the observed band lies in the optically thin segment between $\nu_a$ and $\nu_c$, the closure relations for (a) the Newtonian regime predict temporal and spectral indices of $\alpha = (7p - 5)/6$ and $\beta = (p - 1)/2$, respectively, and (b) for the relativistic regime predict $\alpha = (3p-1)/4$ and  $\beta = (p - 1)/2$, where $F_\nu \propto \nu^{-\beta} t^{-\alpha}$. From the optically thin spectral index derived in Section~\ref{ss_spectra_temp_modelling}, we estimate an average electron energy distribution index of $p = 2.72 \pm 0.08$. Substituting this into the closure relation yields (a) $\alpha = 2.34 \pm 0.09$ in the Newtonian regime and (b) $\alpha = 1.79 \pm 0.06$ in the relativistic regime, which is consistent with the observed light curve decay rate assuming the explosion occurred approximately 1000 days before the first radio detection for wide-angle geometries and approximately 500 days for narow-angle geometries (see Fig.~\ref{Figure:temp_fit}). Alternatively, if the emission is already in the Deep Newtonian regime, the temporal decay would be $\sim t^{-1}$, consistent with a launch date approximately 100 days prior to radio detection \citep{2013ApJ...778..107S}. However, a late launch of $\sim$100 days prior is in tension with the expansion velocity inferred from our equipartition analysis and can be ruled out.

\subsection{The Late-time Radio Lightcurve and Spectrum}

The peaked spectrum of ASKAPJ0055$-$2558 is well fit by the assumed synchrotron self-absorption (SSA) with the ordering $\nu_m < \nu_a < \nu_c$ \citep{2002ApJ...568..820G}. This ordering remains consistent for over 500 days of follow-up multifrequency observations similar to the long-lived evolution observed in GRB970508 \citep{2000ApJ...537..191F} and GRB171205A \citep{2021ApJ...907...60M}. The minimum cooling frequency, $\nu_c$, inferred from equipartition parameters in a Newtonian/relativistic wind model lies in the X-ray band, consistent with its placement above the peak frequency. Our joint SBPL fits yield a temporal evolution of $\nu_a$ with an index of $-0.80^{+0.22}_{-0.21}$. This value is broadly consistent with the theoretical expectation of $-1.24 \pm 0.01$ for a Newtonian, wind-like circumburst medium in the case of a wide-angle jet, given by $-(7p+6)/3(p+4)$. It is also consistent, within uncertainties, with the theoretical expectation of $-1.05 \pm 0.01$ for a relativistic, wind-like circumburst medium in the case of a narrow jet, given by $-3(p+2)/2(p+4)$. Both expectations correspond to an electron power-law index of $p = 2.72 \pm 0.08$ and assume the spectral ordering $\nu_m < \nu_a < \nu_c$ \citep{2000ApJ...537..191F,2013NewAR..57..141G,2021ApJ...907...60M,2002ApJ...568..820G}. 

At 9\,GHz the decline is clearly evident, with the flux decreasing by a factor of $\sim$3 over 600 days. At 0.9\,GHz (see Fig.~\ref{fig:radio_lc}), the decay is much slower and remains poorly constrained due to substantial scatter, though the overall trend is again consistent with a factor of $\sim$1.5 decline over 600 days. The scatter can be naturally explained by refractive interstellar scintillation, since the inferred source size is much smaller than the scattering disk at this frequency (see Section \ref{ss_ISS}).

\subsection{Constraints on the Outflow Launch Time}

Using the radii derived from our equipartition analysis and assuming a constant outflow velocity, a linear back-extrapolation of the late-time expansion implies that the outflow was launched approximately $281 \pm 81$~days before the first radio detection i.e. 2021-04-28, corresponding to the epoch when $R_{\rm eq}\rightarrow 0$ (Fig.~\ref{Figure:Outflow_params}). While this estimate is model-dependent and relies on extrapolating the evolution backward from late times (our monitoring begins $\sim$500~days after the first detection), it provides a physically motivated fiducial launch epoch.

Independent constraints are obtained from standard synchrotron closure relations, which permit launch times in the range of $\sim$500--1000~days prior to the first radio detection. This broad range is further restricted by deep pre-discovery 0.9~GHz non-detections between 2019 and 2021 at  $\sim$0.15~mJy. Between the final non-detection on 2021-05-29 ($<0.15$~mJy) and the first detection on 2022-04-15 (1.30~mJy) at 0.9 GHz, the source brightened by a factor of $\gtrsim 8.7$ in $\lesssim321$~days, which represents an upper limit on the rise rate. We parameterise the rise as a power law, $F_\nu \propto (t - t_0)^{\alpha}$.
To constrain $\alpha$, we construct simple power-law light curves that are required to pass through both the non-detection limit and the first detection, while varying the assumed launch time ($t_0$ = 1, 10, 50, 100, 200, and 500 days prior to the non-detection). As the launch time is placed progressively earlier, the interval between the non-detection and detection constitutes a smaller fraction of the total time since launch, requiring increasingly steep power-law indices to reproduce the observed flux increase while remaining below the pre-discovery limits.
Hence, if the outflow was launched shortly before the 2021-05-29 non-detection, relatively modest rise indices are sufficient (e.g. $\alpha \gtrsim 0.6$ for a launch 10~days earlier, or $\alpha \gtrsim 1.0$ for 50~days earlier). In contrast, launches hundreds of days earlier require much steeper rises, with $\alpha \gtrsim 1.5$ for 100~days, $\alpha \gtrsim 2.1$ for 200~days, and $\alpha \gtrsim 4.0$ for 500~days prior to the non-detection. If the outflow were launched after the last non-detection, assuming the same flux limit the latest possible launch time consistent with the first radio detection would be 70 days earlier, requiring a maximum rise slope of $\alpha \lesssim 0.5$.

As shown in Fig.~\ref{Figure:Lum}, an on-axis relativistic blast-wave launched $\gtrsim 250$~days prior to the first detection should have produced detectable radio emission during the non-detection period. For a standard on-axis GRB or TDE afterglow, launch times $\lesssim 250$~days before the first radio detection would give a slow radio rise which is inconsistent with the rapid rise and decay expected for on-axis jets. Furthermore, if it was launched $\lesssim 250$~days before first radio detection, it should have been fading at a decay rate of $\lesssim1.24$ which is incosistent with the predicted decays rate using the applicable closure relations for $p=2.72$. No GRB localisations overlapped the source position during the 250-day period prior to the first radio detection, therefore, an on-axis GRB interpretation is further discouraged.

In contrast, off-axis relativistic outflows and delayed TDE scenarios naturally accommodate both the observed steep radio rise rate for an earlier launch date and the lack of early-time radio emission. In an off-axis jet scenario, the brightening is primarily driven by the decreasing Lorentz factor, gradually bringing the emission into the observer's line of sight. This geometric effect can produce a characteristically steep rise $\alpha>2$ \citep{2018ApJ...856L..18M}, which would be physically unusual for an on-axis source at such late times but is a hallmark of off-axis evolution. We therefore conclude that while the precise launch epoch remains model dependent, the data favour an outflow launched a few $\gtrsim 250$ days prior to the first radio detection and disfavour scenarios involving an on-axis relativistic jet.

\subsection{Off-axis relativistic jet-like outflow}

\citet{2023MNRAS.522.4565M} extended the equipartition framework to off-axis relativistic jets, wherein the jet initially moves outside our line of sight and later becomes visible as it decelerates and transitions into the Newtonian regime. At the point of this transition, when the jet becomes non-relativistic, the different branches of the equipartition solution converge to a unique solution that satisfies the condition for Newtonian velocity, given by $\beta_{\rm Eq,N}$,

\begin{align}
\beta_{\rm Eq,N} \approx 0.73 
\Bigg[ &
\left( \frac{F_p}{\mathrm{mJy}} \right)^{8/17} 
\left( \frac{d_L}{10^{28}~\mathrm{cm}} \right)^{16/17}
\eta^{35/51} \nonumber \\
& \times 
\left( \frac{\nu_p}{10~\mathrm{GHz}} \right)^{-1} 
(1 + z)^{-8/17} 
\left( \frac{t}{100~\mathrm{d}} \right)^{-1}
\Bigg] \nonumber \\
&\times f_A^{-7/17} f_V^{-1/17} >0.23
\end{align}

Where all parameters are defined in Section~\ref{ss_eqipartition_analysis}. Below 0.23, the relativistic and Newtonian branches split, and the transition should only happen when $\beta_{\rm Eq,N}$ is above this value. For \source\, the inferred $\beta_{\rm Eq,N}$ exceeds the threshold of 0.23 under all assumed collimated geometries, whether time is measured from the first radio detection or shifted 1000 days earlier to approximate the outflow launch (see Fig. \ref{Figure:Outflow_params}). This suggests the source may be transitioning from a collimated off-axis relativistic jet to the Newtonian regime, supporting the applicability of the off-axis jet scenario to this event.

\section{Conclusion}
\label{sec:conc}

Based on all the observational data and analysis of \source\, we draw the following conclusions.

\begin{enumerate}
    \item \source\ is a long-lived, radio transient that exhibited a steep rise in brightness by a factor of at least 20 over 250 days. It has remained detectable in the radio for over 1000 days since its initial discovery.
    
    \item \source\ is very likely ($>98\%$ confidence) associated with the low-mass, star-forming galaxy 2dFGRS TGS143Z140 at a redshift of $z = 0.116$ with prominent HII regions, and lies $\sim$8.7 kpc from the galaxy core in an outer star-forming complex. Deep imaging reveals a potential compact stellar system coincident with the transient position, with a stellar mass of $M_\star \sim 10^{6}$–$10^{8},M_\odot$ and a possible central black hole of $M_{\rm bh} \sim 10^{4}$–$10^{7},M_\odot$. Such an environment is consistent with sites that host GRBs, FBOTs, and off-nuclear IMBH or low-mass SMBH TDEs, supporting both an orphan GRB and a jetted or delayed TDE scenario. If associated to this galaxy, \source\ has a peak radio luminosity of $\rm  \sim10^{29}\:erg\:s^{-1}\:Hz^{-1}$.  

    \item The inferred blast wave velocity, $\beta$, lies in the range $0.2$–$0.9$, placing the transient in the mildly relativistic regime (assuming an explosion date of a day before first radio detection). This holds even if the explosion occurred up to 1000 days before the first radio detection, particularly for more collimated geometries. The corresponding blast wave energy spans $10^{50}$–$10^{53}$ erg, depending on the assumed microphysical parameters. This combination of high energy and mildly relativistic velocities places \source\ in a sparsely populated region of energy–velocity phase space, consistent with the late-time phase of a long GRB afterglow or a delayed or jetted TDE, depending on the outflow geometry.

    \item The electron density evolution with radius, $n_e \propto (R_{\rm eq}/R_s)^{-2.2}$, is consistent with TDEs and with GRBs propagating through a wind-like medium. This scaling supports a scenario in which a jet expands into a low-density, steeply declining circumburst environment, hence, again consistent with a l-GRB afterglow scenario. The inferred densities ($\rm\sim10^{-4}$–$10^{-1}\: cm^{-3}$) are consistent with GRB environments or possibly with the sparse surroundings of IMBHs in globular clusters.
    
    \item The late-time radio light curve and spectra exhibit behaviour consistent with that seen in TDEs and GRBs observed at similarly late epochs. The spectral peak frequency evolves slowly, consistent with either a Newtonian or a relativistic outflow propagating through a wind-like medium. The inferred $\nu_c$ based on the equipartition parameters stays in the X-ray regime, consistent with a $\nu_m,\nu_a<\nu_c$ regime. The 0.9\,GHz light curve continues to exhibit variability during its decay, persisting even 1000 days after the first radio detection. This variability is consistent with the source size remaining smaller than the scattering disk at low frequencies.

     \item Equipartition-based back-extrapolation, synchrotron closure relations, radio non-detections, and rise-time constraints together indicate that the outflow was launched $\gtrsim 250$ days prior to the first radio detection. These considerations disfavour an on-axis event and instead favour an interpretation in which \source\ represents late-time emission from either an orphan GRB afterglow or an off-axis or delayed TDE. An off-axis geometry is also supported by the inferred Newtonian velocity, $\beta_{\rm eq,N} > 0.23$, across a range of jet and ejecta configurations. 
    
\end{enumerate}

In conclusion, the properties of \source\ are most consistent with the late-time detection of an orphan l-GRB afterglow or an off-nuclear TDE involving an IMBH. If ASKAP J0055–2558 is indeed an orphan afterglow detected in the radio, it would represent only the second such discovery to date \citep{law_discovery_2018,2019ApJ...876L..14M,mooley_late-time_2022}, and arguably the most compelling candidate of its kind. Alternatively, if it is an off-nuclear TDE associated with an IMBH, it would mark the first such event discovered exclusively via radio observations.

At such late times, when the jet has entered its Newtonian phase, distinguishing between progenitor classes becomes intrinsically difficult. In our case, the UV-bright host complicated the isolation of any long-lived accretion-disk emission expected from a TDE \citep{gezari_tidal_2021}, whereas in a UV-quiet host, such a signal (or its absence) would offer a clearer diagnostic. Early X-ray observations could have been decisive: off-axis GRB afterglows are typically faint in X-rays at early times, while jetted TDEs can produce luminous, long-lived X-ray emission  \citep{zauderer_birth_2011,levan_extremely_2011}. Similarly, uninterrupted optical photometric monitoring would have constrained the rise time, helping better panchromatic modelling, but there was a $~$200 day gap in ZTF monitoring coincident with the last radio non-detection (2021-01-15 - 2021-08-09) and the first radio detection (2021-12-2 - 2022-08-12). A prompt optical spectrum of the transient would have provided an additional, independent diagnostic through line features or continuum shape \citep{2025arXiv250803807S}. Future discoveries of this kind will benefit most from rapid multi-wavelength follow-up, particularly early X-ray/UV coverage, dense radio monitoring, and timely optical spectroscopy to break the orphan-afterglow versus off-nuclear TDE degeneracy.

\balance

\section{Acknowledgements}

We are grateful to the reviewer for their insightful and helpful comments, which have strengthened the manuscript. We thank Marcus Lower for helping with the Parkes UWL observations and Themiya Nanayakara for a discussion on optical analysis techniques. AG is supported by the Australian Government Research Training
Program Scholarship. This research was conducted
with support from the Australian Research Council Centre of Excellence for Gravitational Wave Discovery (OzGrav), project number CE230100016. AM is supported by the ARC Discovery Early Research Award (DE230100055).
This scientific work uses data obtained from Inyarrimanha Ilgari Bundara / the CSIRO's Murchison Radio-astronomy Observatory. We acknowledge the Wajarri Yamaji People as the Traditional Owners and native title holders of the Observatory site. CSIRO’s ASKAP radio telescope is part of the \href{https://ror.org/05qajvd42}{Australia Telescope National Facility}. Operation of ASKAP is funded by the Australian Government with support from the National Collaborative Research Infrastructure Strategy. ASKAP uses the resources of the Pawsey Supercomputing Research Centre. Establishment of ASKAP, Inyarrimanha Ilgari Bundara, the CSIRO Murchison Radio-astronomy Observatory and the Pawsey Supercomputing Research Centre are initiatives of the Australian Government, with support from the Government of Western Australia and the Science and Industry Endowment Fund.
We acknowledge the Gomeroi people as the traditional owners of the Australia Telescope Compact Array (ATCA) observatory site. 
The ATCA is part of the Australia Telescope National Facility, which is funded by the Australian Government for operation as a National Facility managed by CSIRO. 
Jeff Cooke acknowledges funding from the Australian Research Council Discovery Project DP200102102.
This work is (partly) based on data obtained through Swinburne.
Keck program 2024B\_W377; Keck Observatory is a private 501(c)3 non-profit organisation operated as a scientific partnership among the California Institute of Technology, the University of California, and the National Aeronautics and Space Administration. The Observatory was made possible by the generous financial support of the W. M. Keck Foundation. The authors wish to recognise and acknowledge the very significant cultural role and reverence that the summit of Maunakea has always had within the Native Hawaiian community. We are most fortunate to have the opportunity to conduct observations from this mountain. AI tools (ChatGPT-4) were used for language editing and clarity. AG takes full responsibility for the content.

\textit{Software:} \texttt{VAST Tools} \citep{vast_tools}, \texttt{MIRIAD} \citep{1995ASPC...77..433S}, \texttt{CASA} \citep{2007ASPC..376..127M}, \texttt{XRTPIPELINE} \citep{Evans:2009}, \texttt{Scarlet} \citep{Melchior2018}, \texttt{LIME} \citep{2024A&A...688A..69F}, \texttt{IRAF} \citep{Tody86}, \texttt{SExtractor} \citep{Bertin1996}, \texttt{SWARP} \citep{2002ASPC..281..228B}, \texttt{MARZ}\citep{Hinton:2016}

\appendix

\section{Radio Observational Data and Fits}
\label{app:radio}

\begin{center}
\begin{longtable*}{ccccccc}
\caption{The radio data used in this work. All upper limits are $5\sigma$ and errors are $1\sigma$.\label{Table:Radio_obs}}\\
\hline
\hline
Epoch &
  \begin{tabular}[c]{@{}c@{}}Frequency \\ (GHz)\end{tabular} &
  \begin{tabular}[c]{@{}c@{}}Flux Density\\ (mJy)\end{tabular} &
  \begin{tabular}[c]{@{}c@{}}Flux Density\\ Error (mJy)\end{tabular} &
  \begin{tabular}[c]{@{}c@{}}ISS Error \\ (mJy)\end{tabular} &
  \begin{tabular}[c]{@{}c@{}}RMS \\ (mJy)\end{tabular} &
  \begin{tabular}[c]{@{}c@{}}Telescope,\\ Survey, Project\end{tabular} \\ \hline
2019-04-26 & 0.887  & \textless{}1.06 & -    & -    & 0.21 & ASKAP, RACS   \\
2019-07-06 & 3.0    & \textless{}0.68 & -    & -    & 0.14 & VLA, VLASS    \\
2019-08-16 & 0.943  & \textless{}0.15 & -    & -    & 0.03 & ASKAP, AS111  \\
2019-08-23 & 0.943  & \textless{}0.16 & -    & -    & 0.03 & ASKAP, AS111  \\
2019-09-16 & 0.943  & \textless{}0.15 & -    & -    & 0.03 & ASKAP, AS111  \\
2019-11-07 & 0.943  & \textless{}0.15 & -    & -    & 0.03 & ASKAP, AS111  \\
2020-04-03 & 0.943  & \textless{}0.12 & -    & -    & 0.02 & ASKAP, AS111  \\
2020-04-29 & 0.943  & \textless{}0.17 & -    & -    & 0.03 & ASKAP, AS111  \\
2020-07-03 & 0.943  & \textless{}0.13 & -    & -    & 0.03 & ASKAP, AS111  \\
2020-11-28 & 0.943  & \textless{}0.15 & -    & -    & 0.03 & ASKAP, AS111  \\
2020-11-29 & 0.943  & \textless{}0.17 & -    & -    & 0.03 & ASKAP, AS111  \\
2021-01-06 & 1.3675 & \textless{}0.78 & -    & -    & 0.16 & ASKAP, RACS   \\
2021-05-29 & 0.943  & \textless{}0.15 & -    & -    & 0.03 & ASKAP, AS111  \\
2022-02-04 & 3.0    & 2.91            & 0.29 & 1.73 & 0.14 & VLA, VLASS    \\
2022-04-15 & 0.887  & 1.30            & 0.23 & 0.39 & 0.13 & ASKAP, RACS   \\
2023-05-04 & 1.332  & \textless{}1.83 & -    & -    & 0.37 & ATCA, C3363   \\
2023-05-04 & 1.844  & 1.97            & 0.15 & 0.85 & 0.08 & ATCA, C3363   \\
2023-05-04 & 2.1    & 1.99            & 0.11 & 0.97 & 0.08 & ATCA, C3363   \\
2023-05-04 & 2.356  & 1.92            & 0.11 & 0.98 & 0.07 & ATCA, C3363   \\
2023-05-04 & 2.868  & 1.97            & 0.12 & 1.17 & 0.09 & ATCA, C3363   \\
2023-05-04 & 5.5    & 1.37            & 0.02 & 0.06 & 0.02 & ATCA, C3363   \\
2023-05-04 & 9.0    & 0.90            & 0.02 & 0.01 & 0.01 & ATCA, C3363   \\
2023-06-08 & 1.332  & 2.22            & 0.13 & 0.83 & 0.12 & ATCA, C3363   \\
2023-06-08 & 1.844  & 2.33            & 0.07 & 1.01 & 0.06 & ATCA, C3363   \\
2023-06-08 & 2.1    & 2.14            & 0.06 & 1.04 & 0.07 & ATCA, C3363   \\
2023-06-08 & 2.356  & 2.13            & 0.06 & 1.10 & 0.04 & ATCA, C3363   \\
2023-06-08 & 2.868  & 2.00            & 0.10 & 1.20 & 0.07 & ATCA, C3363   \\
2023-06-08 & 5.5    & 1.34            & 0.02 & 0.06 & 0.02 & ATCA, C3363   \\
2023-06-08 & 9.0    & 0.93            & 0.02 & 0.01 & 0.02 & ATCA, C3363   \\
2023-07-07 & 1.28   & 1.97            & 0.03 & 0.73 & 0.38 & MeerKat       \\
2023-08-27 & 1.332  & 2.49            & 0.14 & 0.92 & 0.17 & ATCA, C3363   \\
2023-08-27 & 1.844  & 2.63            & 0.10 & 1.14 & 0.08 & ATCA, C3363   \\
2023-08-27 & 2.1    & 2.33            & 0.05 & 1.14 & 0.08 & ATCA, C3363   \\
2023-08-27 & 2.356  & 2.38            & 0.10 & 1.22 & 0.08 & ATCA, C3363   \\
2023-08-27 & 2.868  & 2.00            & 0.13 & 1.19 & 0.09 & ATCA, C3363   \\
2023-08-27 & 5.5    & 1.21            & 0.03 & 0.06 & 0.02 & ATCA, C3363   \\
2023-08-27 & 9.0    & 0.80            & 0.02 & 0.01 & 0.02 & ATCA, C3363   \\
2023-10-15 & 0.887  & 1.66            & 0.17 & 0.50 & 0.18 & ASKAP, VAST   \\
2023-11-29 & 1.332  & 2.31            & 0.18 & 0.86 & 0.13 & ATCA, C3363   \\
2023-11-29 & 1.844  & 2.07            & 0.07 & 0.90 & 0.06 & ATCA, C3363   \\
2023-11-29 & 2.1    & 2.05            & 0.05 & 1.00 & 0.04 & ATCA, C3363   \\
2023-11-29 & 2.356  & 1.98            & 0.06 & 1.02 & 0.04 & ATCA, C3363   \\
2023-11-29 & 2.868  & 1.73            & 0.07 & 1.03 & 0.05 & ATCA, C3363   \\
2023-11-29 & 5.5    & 1.13            & 0.02 & 0.05 & 0.02 & ATCA, C3363   \\
2023-11-29 & 9.0    & 0.74            & 0.02 & 0.01 & 0.01 & ATCA, C3363   \\
2024-01-02 & 0.943  & 2.91            & 0.27 & 0.88 & 0.23 & ASKAP, RACS   \\
2024-01-06 & 0.943  & 2.50            & 0.13 & 0.75 & 0.17 & ASKAP, RACS   \\
2024-03-10 & 1.332  & 1.91            & 0.12 & 0.71 & 0.13 & ATCA, C3363   \\
2024-03-10 & 1.844  & 1.76            & 0.07 & 0.76 & 0.07 & ATCA, C3363   \\
2024-03-10 & 2.1    & 1.70            & 0.04 & 0.83 & 0.06 & ATCA, C3363   \\
2024-03-10 & 2.356  & 1.69            & 0.06 & 0.87 & 0.05 & ATCA, C3363   \\
2024-03-10 & 2.868  & 1.65            & 0.08 & 0.99 & 0.06 & ATCA, C3363   \\
2024-03-10 & 5.5    & 0.93            & 0.03 & 0.04 & 0.02 & ATCA, C3363   \\
2024-03-10 & 9.0    & 0.56            & 0.02 & 0.00 & 0.02 & ATCA, C3363   \\
2024-04-05 & 0.887  & 2.30            & 0.15 & 0.69 & 0.16 & ASKAP, VAST   \\
2024-05-01 & 0.67   & 1.38           & 0.29 & 0.36 & 0.16 & GMRT, 46\_054 \\
2024-05-03 & 1.27   & 1.68            & 0.13 & 0.62 & 0.13 & GMRT, 46\_054 \\
2024-05-05 & 0.4    & 0.87            & 0.05 & 0.17 & 0.05 & GMRT, 46\_054 \\
2024-06-01 & 0.887  & 1.94            & 0.16 & 0.59 & 0.19 & ASKAP, , VAST \\
2024-06-22 & 1.27   & 1.72 & 0.10 & 0.64 & 0.14 & GMRT, 46\_054  \\
2024-06-21 & 0.67   & 0.96            & 0.06 & 0.25 & 0.13 & GMRT, 46\_054 \\
2024-06-22 & 0.4    & \textless{}1.05 & -    & -    & 0.21 & GMRT, 46\_054 \\
2024-06-24 & 1.332  & 1.80            & 0.33 & 0.67 & 0.21 & ATCA, C3363   \\
2024-06-24 & 1.844  & 1.74            & 0.14 & 0.75 & 0.09 & ATCA, C3363   \\
2024-06-24 & 2.1    & 1.43            & 0.06 & 0.70 & 0.06 & ATCA, C3363   \\
2024-06-24 & 2.356  & 1.57            & 0.07 & 0.81 & 0.04 & ATCA, C3363   \\
2024-06-24 & 2.868  & 1.36            & 0.07 & 0.81 & 0.05 & ATCA, C3363   \\
2024-06-24 & 5.5    & 0.86            & 0.02 & 0.04 & 0.02 & ATCA, C3363   \\
2024-06-24 & 9.0    & 0.56            & 0.02 & 0.00 & 0.01 & ATCA, C3363   \\
2024-08-02 & 0.887  & 2.44            & 0.15 & 0.74 & 0.17 & ASKAP, VAST   \\
2024-09-24 & 1.332  & 1.66            & 0.18 & 0.62 & 0.14 & ATCA, C3363   \\
2024-09-24 & 1.844  & 1.70            & 0.09 & 0.74 & 0.06 & ATCA, C3363   \\
2024-09-24 & 2.1    & 1.61            & 0.05 & 0.78 & 0.06 & ATCA, C3363   \\
2024-09-24 & 2.356  & 1.55            & 0.06 & 0.80 & 0.04 & ATCA, C3363   \\
2024-09-24 & 2.868  & 1.27            & 0.07 & 0.76 & 0.05 & ATCA, C3363   \\
2024-09-24 & 5.5    & 0.81            & 0.03 & 0.04 & 0.02 & ATCA, C3363   \\
2024-09-24 & 9.0    & 0.54            & 0.02 & 0.00 & 0.01 & ATCA, C3363   \\
2024-09-28 & 0.887  & 1.81            & 0.14 & 0.55 & 0.17 & ASKAP, VAST   \\
2024-09-28 & 3.0    & 1.02            & 0.15 & 0.61 & 0.15 & VLA, VLASS    \\
2024-12-26 & 0.4    & \textless{}3.00 & -    & -    & 0.60 & GMRT, 46\_054 \\
2024-12-26 & 0.67   & 1.54            & 0.17 & 0.40 & 0.08 & GMRT, 46\_054 \\
2024-12-26 & 1.332  & 1.75            & 0.12 & 0.65 & 0.10 & ATCA, C3363   \\
2024-12-26 & 1.844  & 1.69            & 0.06 & 0.73 & 0.05 & ATCA, C3363   \\
2024-12-26 & 2.1    & 1.42            & 0.04 & 0.69 & 0.06 & ATCA, C3363   \\
2024-12-26 & 2.356  & 1.30            & 0.06 & 0.67 & 0.04 & ATCA, C3363   \\
2024-12-26 & 2.868  & 1.08            & 0.07 & 0.64 & 0.05 & ATCA, C3363   \\
2024-12-26 & 5.5    & 0.74            & 0.02 & 0.03 & 0.01 & ATCA, C3363   \\
2024-12-26 & 9.0    & 0.42            & 0.02 & 0.00 & 0.01 & ATCA, C3363   \\
2025-02-22 & 0.887  & 2.01            & 0.18 & 0.61 & 0.18 & ASKAP, VAST   \\
2025-03-20 & 1.332  & 1.82            & 0.11 & 0.67 & 0.11 & ATCA, C3363   \\
2025-03-20 & 1.844  & 1.63            & 0.06 & 0.70 & 0.05 & ATCA, C3363   \\
2025-03-20 & 2.1    & 1.45            & 0.04 & 0.71 & 0.06 & ATCA, C3363   \\
2025-03-20 & 2.356  & 1.19            & 0.06 & 0.61 & 0.04 & ATCA, C3363   \\
2025-03-20 & 2.868  & 1.30            & 0.09 & 0.78 & 0.06 & ATCA, C3363   \\
2025-03-20 & 5.5    & 0.61            & 0.03 & 0.03 & 0.02 & ATCA, C3363   \\
2025-03-20 & 9.0    & 0.31            & 0.02 & 0.00 & 0.02 & ATCA, C3363   \\
2025-04-21 & 0.887  & 1.84            & 0.17 & 0.56 & 0.18 & ASKAP, VAST \\
\hline
\end{longtable*}
\end{center}

\begin{figure*}[b]
    \centering
    \includegraphics[width=0.85\textwidth]{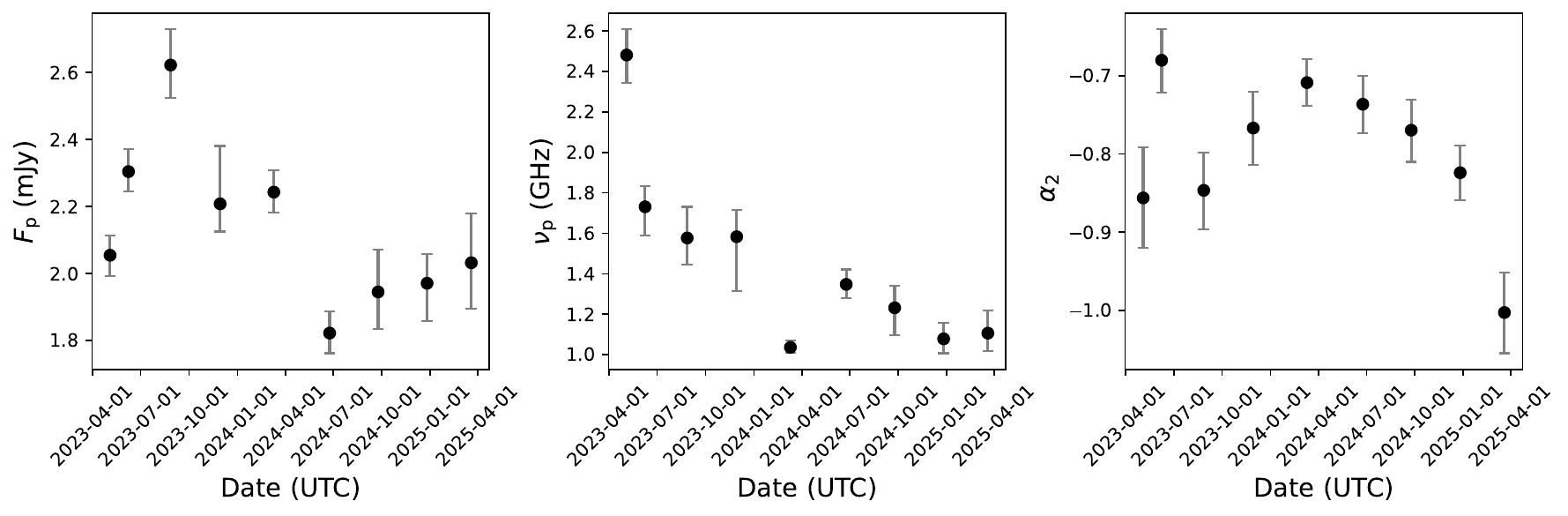}
    \\
    \includegraphics[width=0.85\textwidth]{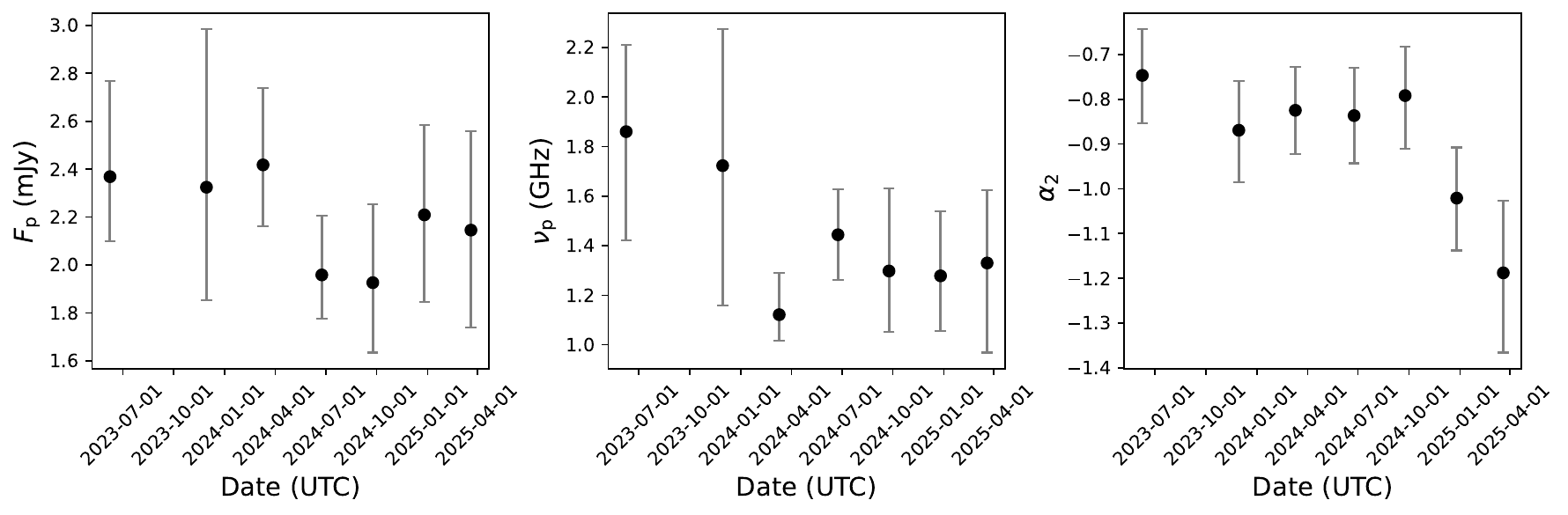}
    \\
    \includegraphics[width=0.85\textwidth]{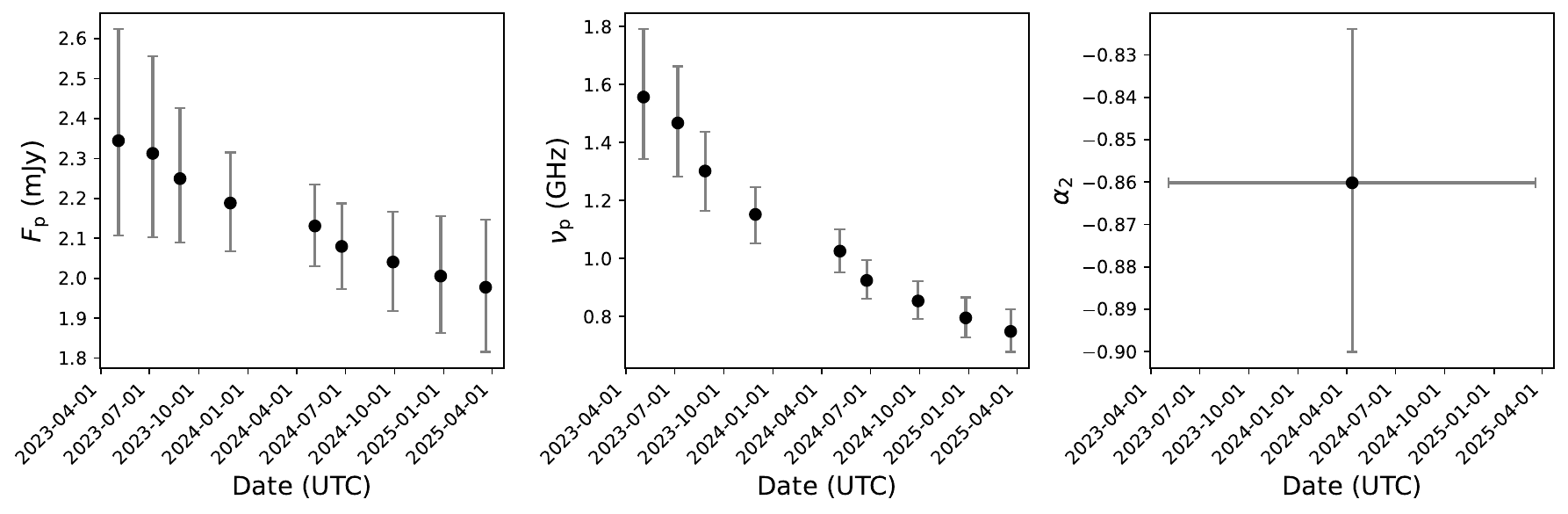}
  \\
    \caption{Temporal evolution of the SBPL spectral fit parameters for ASKAP J0055–2558 across nine radio epochs. The top panel shows the results of the initial fits without including interstellar scintillation (ISS) and without a joint fit. The middle panel shows the results with ISS included and with individual epoch SBPL fits applied. The bottom panel shows the results with ISS included and with a joint fit applied. For the joint fit, the peak flux and frequency were derived for each epoch from the global fit. In each panel, the parameters are displayed as a function of observation date: peak flux density ($F_p$, right), peak frequency ($\nu_p$, middle), and optically thin spectral index ($\alpha_2$, left). Error bars represent 1$\sigma$ uncertainties from the MCMC fitting.}
    \label{fig:SBPL_params}
\end{figure*}

\FloatBarrier
\begin{deluxetable}{ccccc}[]
\tablecaption{Best-Fit SBPL Parameters for ASKAP~J0055$-$2558 Radio Spectra \label{tab:spectra_fits}}
\tablehead{
\colhead{Observation Date} & 
\colhead{$F_p$} & 
\colhead{$\nu_p$} & 
\colhead{$\alpha_2$} & 
\colhead{$p$} \\
\colhead{(UTC)} & 
\colhead{(mJy)} & 
\colhead{(GHz)} & 
\colhead{} & 
\colhead{}
}
\startdata
2023-05-04 & $2.34^{+0.28}_{-0.24}$ & $1.56^{+0.23}_{-0.21}$ & &  \\
2023-07-07 & $2.31^{+0.24}_{-0.21}$ & $1.47^{+0.20}_{-0.19}$ & &  \\
2023-08-27 & $2.25^{+0.18}_{-0.16}$ & $1.30^{+0.14}_{-0.14}$ & &  \\
2023-11-29 & $2.19^{+0.13}_{-0.12}$ & $1.15^{+0.10}_{-0.10}$ & &  \\
2024-05-05 & $2.13^{+0.10}_{-0.10}$ & $1.03^{+0.08}_{-0.07}$ & $-0.86^{+0.04}_{-0.04}$ & $2.72^{+0.07}_{-0.04}$ \\
2024-06-24 & $2.08^{+0.11}_{-0.11}$ & $0.92^{+0.07}_{-0.06}$ & &  \\
2024-09-28 & $2.04^{+0.13}_{-0.12}$ & $0.85^{+0.07}_{-0.06}$ & &  \\
2024-12-26 & $2.01^{+0.15}_{-0.14}$ & $0.79^{+0.07}_{-0.07}$ & &  \\
2025-03-20 & $1.98^{+0.17}_{-0.16}$ & $0.75^{+0.08}_{-0.07}$ & &  \\
 \hline
\enddata
\tablecomments{Best-fit parameters from the joint fit smoothly broken power-law (SBPL) model applied to the radio spectra of ASKAP~J0055$-$2558 across nine epochs simultaneously. Listed parameters are the derived peak flux density ($F_p$), peak frequency ($\nu_p$), high-frequency spectral index ($\alpha_2$), and electron power-law index ($p$). The joint fit yields $\chi^2_{\rm red} = 0.89$ with $\sim 74$ degrees of freedom. Uncertainties are 1$\sigma$ credible intervals from MCMC fitting.}
\end{deluxetable}


\section{Extended Multiwavelength Analysis}
\label{app:mw}

\begin{figure*}[t]
    \centering
    \begin{minipage}{0.49\textwidth}
        \centering
        \includegraphics[width=\linewidth]{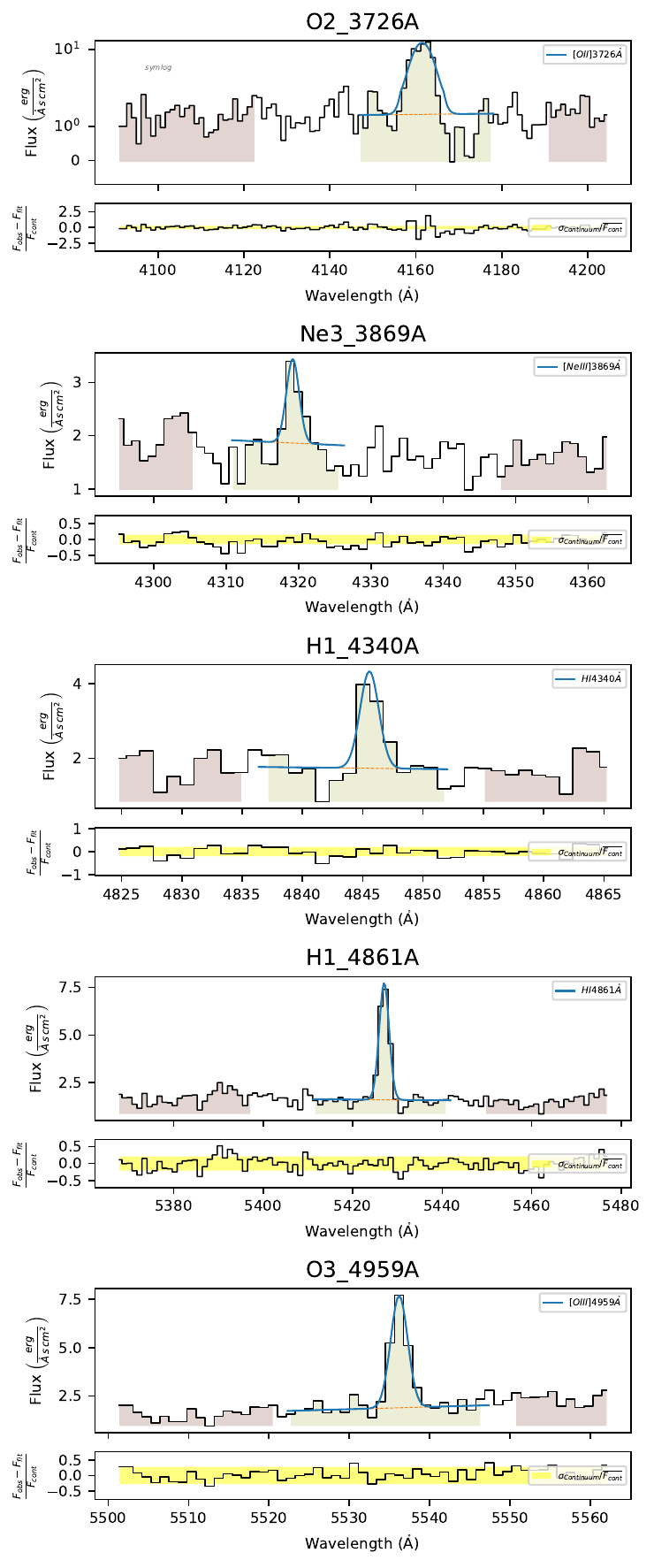}
    \end{minipage}\hfill
    \begin{minipage}{0.49\textwidth}
        \centering
        \includegraphics[width=\linewidth]{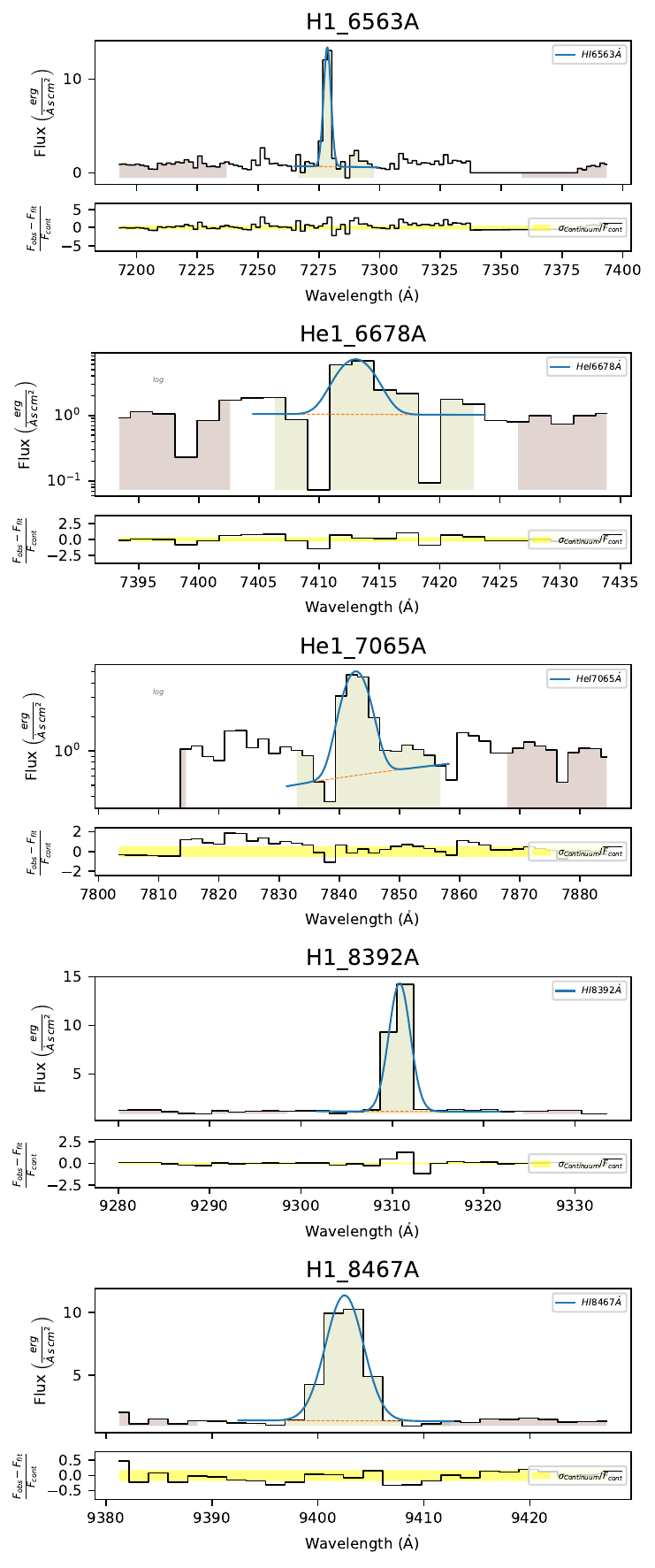}
    \end{minipage}
    \caption{Spectral fitting results obtained using the LIME package for the galaxy core, showing the best-fit model (blue solid line) overlaid on the observed spectrum. The continuum (bottom panels) and emission-line (top panels) components are indicated separately.}
    \label{fig:optical-fits}
\end{figure*}

\begin{figure*}[t]
    \centering
    \begin{minipage}{1\textwidth}
        \centering
        \includegraphics[width=\linewidth,trim={0 4.25cm 0 0},clip]{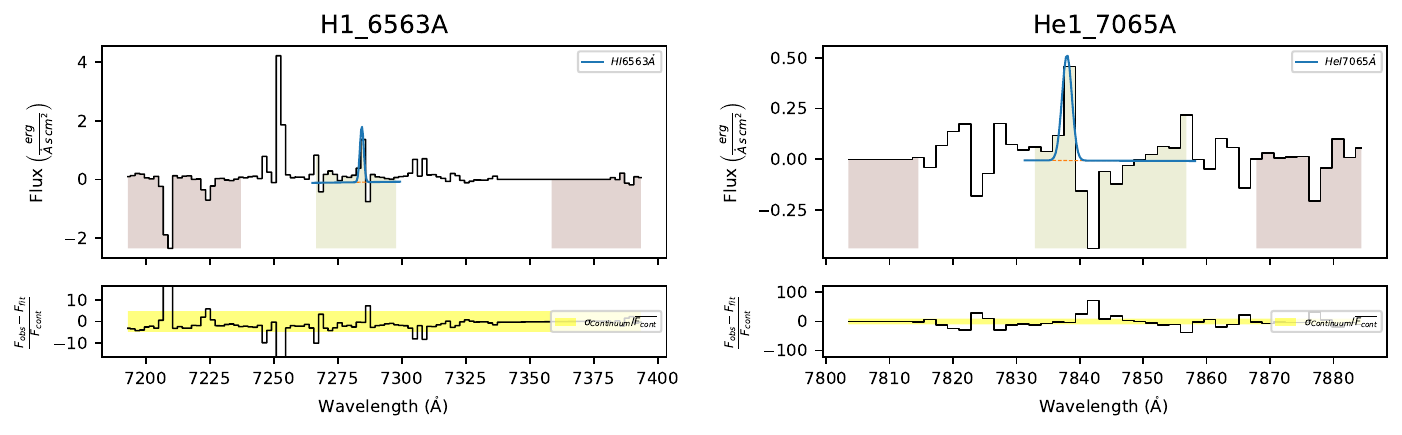}
    \end{minipage}
    \\
    \vspace{-0.25cm}
    \begin{minipage}{1\textwidth}
        \centering
        \hspace{-5pt}\includegraphics[width=\linewidth]{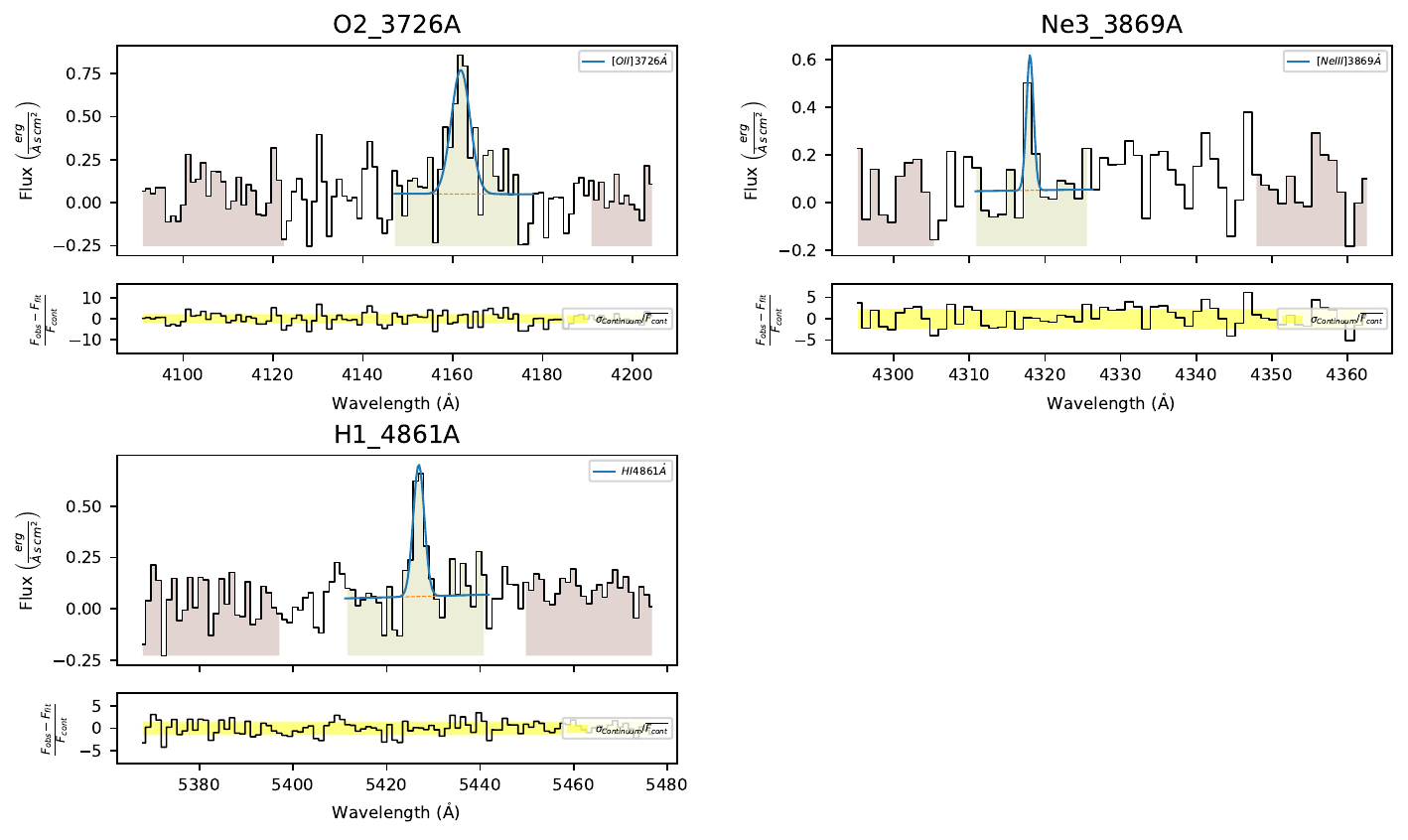}
    \end{minipage}
    \caption{Spectral fitting results obtained using the LIME package at the transient position, showing the best-fit model (blue solid line) overlaid on the observed spectrum. The continuum (bottom panels) and emission-line (top panels) components are indicated separately.}
    \label{fig:optical-fits_trans_pos_fit}
\end{figure*}

\begin{figure*}[htbp]
    \centering
        \includegraphics[width=\linewidth]{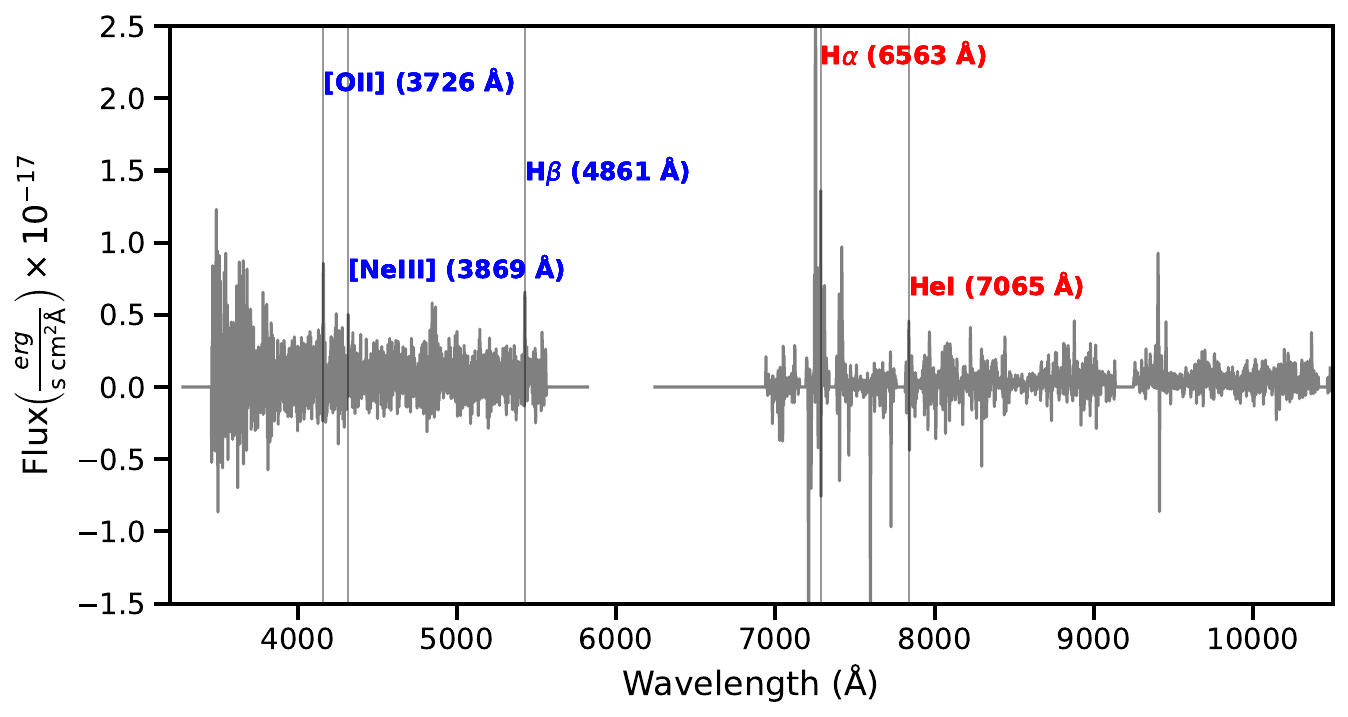}
    \caption{Spectra obtained with Keck KCWI and KCRM at the transient position, with emission lines fitted using the \texttt{LIME} package. The signal-to-noise ratio, as defined by \citet{1994A&A...287..676R} for each of the fitted lines, is above 5$\sigma$. The fitting results are given in the Appendix \ref{app:mw} Fig. \ref{fig:optical-fits_trans_pos_fit}.}
    \label{fig:optical-fits_transient_pos}
\end{figure*}

\subsection{GRB Chance-coincidence estimation}

To test whether spatially coincident GRBs could be genuine counterparts, we estimated chance-alignment probabilities via Monte Carlo simulations. Using the first radio detection (2022 February 4 UTC) as the reference date, we considered GRBs within time windows of $Y$ years prior and generated $N=300\times Y$ synthetic bursts uniformly distributed over the sky to match the observed GRB rate. Repeating this 500 times per window, we calculated the mean hit fraction $f_{\mathrm{hit}}$ (probability of chance coincidence). For $Y=5$ yr (2 GRBs) $f_{\mathrm{hit}}\approx1.1\%$, for $Y=10$ yr (3 GRBs) $\approx2.5\%$, for $Y=20$ yr (12 GRBs) $\approx23.8\%$, and for $Y=30$ yr (54 GRBs) $\approx55.5\%$. Thus, chance-coincidence probabilities are low for $\lesssim10$ yr windows but become substantial over $\sim30$ yr.

\subsection{Host Galaxy Chance Coincidence Estimation}
\label{app:hg_chance_allignment}

To estimate the chance alignment of the galaxy offset from \source\ we used the method described in \citet{2002AJ....123.1111B} and the code from \texttt{astropath} \citep{2021ApJ...911...95A}. For each galaxy of apparent magnitude $r_{\mathrm{mag}}$, the number density of field galaxies on the sky, $n_{\mathrm{den}}$, is estimated from the empirical relation of \citet{2002AJ....123.1111B} available in \texttt{astropath}. The effective search radius is conservatively defined as
\begin{equation}
\theta_{\mathrm{eff}} = \sqrt{4\sigma_R^2 + 4r_{1/2}^2 + \mathrm{sep}^2},
\end{equation}
where $\sigma_R$ is the $1\sigma$ localisation error of the transient, $r_{1/2}$ is the galaxy half-light radius, and $\mathrm{sep}$ is the measured angular offset between the transient and the galaxy centre. The expected number of unrelated galaxies within the search area is $\bar{N} = \pi\,\theta_{\mathrm{eff}}^2\,n_{\mathrm{den}}$ and the corresponding probability of a chance alignment is
\begin{equation}
P_{\mathrm{chance}} = 1 - e^{-\bar{N}}.
\end{equation}

All these calculations are directly run within \texttt{astropath}, which assumes galaxies follow a Poisson distribution on the sky. The returned values are $P_{\mathrm{chance}}$, representing the probability that a galaxy of magnitude $\leq r_{\mathrm{mag}}$ would fall within the observed offset by chance. This probability of chance coincidence for our transient in relation to the epotential host galaxy has been quoted in Section \ref{sec:host_galaxy}.

\subsection{Host Galaxy Physical Parameters Calculations}
\label{Appendix:Gal_calcs}

The host galaxy is clearly detected in the DESI Legacy Imaging Surveys in the g, r, and i optical bands and in the near-infrared J band from Keck imaging. We therefore derived its stellar population properties by fitting stellar population synthesis (SPS) templates to the available broadband photometry. The fitting was performed with the \texttt{Bagpipes} code \citep{2018MNRAS.480.4379C}, which models galaxy spectral energy distributions (SEDs) using parametric star-formation histories and a Bayesian framework. The input photometry included GALEX FUV and NUV fluxes together with the DESI g, r, i, z and Keck J measurements.  From the resulting posterior distributions, we extracted the galaxy’s stellar mass, star-formation rate, and metallicity parameters quoted in Section \ref{sec:host_galaxy}.




From the observed Balmer decrement (H$\alpha$/H$\beta = 4.00$ vs.\ the intrinsic 2.86; \citealt{1989agna.book.....O}), we derive color excesses of $E(B-V)=0.27$ (Calzetti law; \citealt{2000ApJ...533..682C}) corresponding to $A_V \approx 1.1$ mag. Line fluxes are then corrected for dust attenuation using the wavelength-dependent reddening law \citep{1994ApJ...429..582C,1997AIPC..408..403C}. The [O\,\textsc{iii}]/[O\,\textsc{ii}] ratio is a widely used probe of ionization conditions, where the total [O\,\textsc{iii}] flux is inferred from the 4959\,\AA\ line assuming the theoretical 5007/4959 ratio of 2.98 \citep{2007MNRAS.374.1181D}. The [O\,\textsc{ii}] $\lambda3727$ doublet traces singly ionised oxygen (O$^{+}$), arising in lower-ionisation zones, and thus provides a complementary measure to the higher-ionisation [O\,\textsc{iii}] lines. For the main galaxy, dust-corrected fluxes give [O\,\textsc{iii}]/[O\,\textsc{ii}] ratios of 0.92 (Calzetti law consistent with a moderately ionised interstellar medium.

\begin{figure*}[b]
    \centering
        \includegraphics[width=0.9\linewidth]{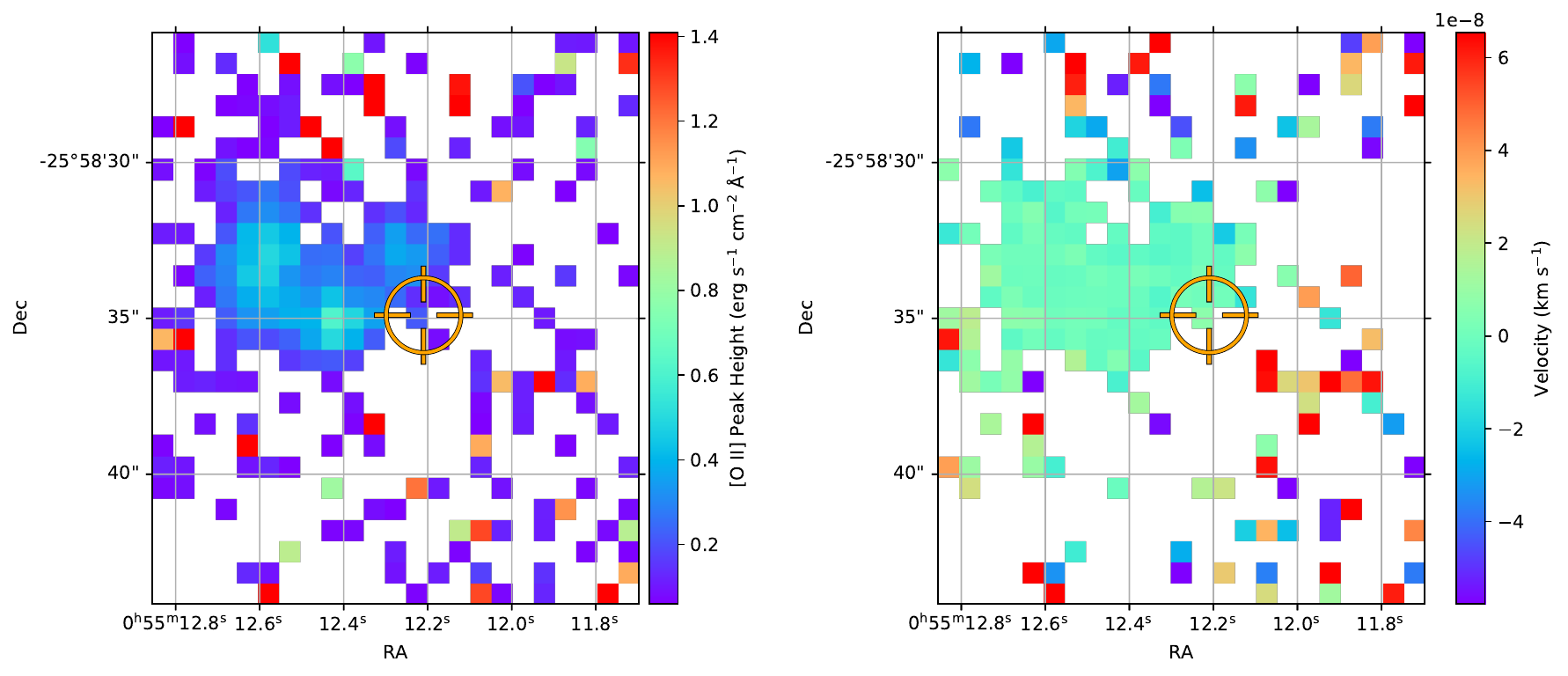}
    \caption{Velocity map of the ionized gas traced by the \ion{O}{2} [3726] emission line. The plot on the left indicates the line intensity (height of the fitted emission line), while the plot on the right shows the relative line-of-sight velocity. Only spaxels with reliable \ion{O}{2} detections are displayed. QFitsView and DPUserFunction (\url{https://github.com/MarkDurre/DPUserFunctions.git})  were used for making the map. The aggregate source position of the transient with uncertainty derived from ATCA and GMRT observations is also overlaid.}
    \label{fig:velmap}
\end{figure*}

\begin{deluxetable}{lccccccc}[h]
\label{Table:Optical_Spectra_fits}
\tablecaption{Gaussian Fit Parameters for Emission Lines.}
\tablehead{
\colhead{Line} &
\colhead{Rest Wavelength} & 
\colhead{Center} & 
\colhead{Amplitude} & 
\colhead{Gaussian Width} & 
\colhead{Integrated Flux} & 
\colhead{Reduced $\chi^2$} & \colhead{DOF}\\
\colhead{} &
\colhead{($\rm \AA$)} &
\colhead{($\rm \AA$)} &
\colhead{($10^{-17}\,\mathrm{erg\,s^{-1}\,cm^{-2}\,\AA^{-1}}$)} &
\colhead{($\rm \AA$)} &
\colhead{($10^{-17}\,\mathrm{erg\,s^{-1}\,cm^{-2}}$)} &
\colhead{} &
\colhead{}
} 

\startdata
\hline
\multicolumn{8}{c}{Galaxy Nucleus} \\ \hline
H$\alpha$ (6563 $\rm \AA$)   & 6563.0 & 7278.46 $\pm$ 0.08 & 12.75 $\pm$ 0.63  & 1.51 $\pm$ 0.09 & 59.51 $\pm$ 2.98 & 2.60 & 63\\
\ion{He}{1} (6678 $\rm \AA$)         & 6678.0 & 7413.02 $\pm$ 0.17 & 6.16 $\pm$ 0.71   & 1.37 $\pm$ 0.19 & 23.77 $\pm$ 2.30 & 2.63  & 21\\
\ion{He}{1} (7065 $\rm \AA$)         & 7065.0 & 7842.75 $\pm$ 0.15 & 4.46 $\pm$ 0.29   & 2.01 $\pm$ 0.15 & 25.44 $\pm$ 2.23 & 1.08 & 31\\
\ion{H}{1} (8392 $\rm \AA$)          & 8392.0 & 9310.72 $\pm$ 0.46 & 13.23 $\pm$ 0.59 & 1.17 $\pm$ 0.07 & 40.46 $\pm$ 0.84 & 9.96 & 28\\
\ion{H}{1} (8467 $\rm \AA$)          & 8467.0 & 9402.52 $\pm$ 0.05 & 10.08 $\pm$ 0.24  & 1.79 $\pm$ 0.05 & 41.90 $\pm$ 1.48 & 1.21 & 23\\
\ion{O}{2} (3726 $\rm \AA$)   & 3726.0 & 4161.44 $\pm$ 0.10 & 11.49 $\pm$ 0.49  & 2.11 $\pm$ 0.10 & 52.64 $\pm$ 2.20 & 4.31 & 70\\
\ion{Ne}{3} (3869 $\rm \AA$) & 3869.0 & 4319.19 $\pm$ 0.21 & 1.58 $\pm$ 0.35   & 0.88 $\pm$ 0.23 & -0.16 $\pm$ 1.20 & 1.35 & 38\\
H$\gamma$ (4340 $\rm \AA$)   & 4340.0 & 4845.56 $\pm$ 0.14 & 2.60 $\pm$ 0.52   & 0.78 $\pm$ 0.20 & 3.23 $\pm$ 1.51  & 1.15 & 34\\
H$\beta$ (4861 $\rm \AA$)    & 4861.0 & 5427.03 $\pm$ 0.06 & 6.14 $\pm$ 0.28   & 1.12 $\pm$ 0.06 & 14.90 $\pm$ 1.92 & 0.87 & 79\\
\ion{O}{3} (4959 $\rm \AA$)  & 4959.0 & 5536.20 $\pm$ 0.09 & 5.79 $\pm$ 0.44   & 1.06 $\pm$ 0.09 & 17.20 $\pm$ 2.67 & 0.78
& 51\\
\hline
\multicolumn{8}{c}{Transient Position} \\ \hline
H$\alpha$ (6563 $\rm \AA$) & 6563.0 & $7284.34\pm2.36$ & $1.90\pm8.43$ & $0.75\pm2.55$ & $8.34\pm3.57$ & 0.91 & 63\\
\ion{He}{1} (7065 $\rm \AA$) & 7065.0 & $7837.97\pm1.38$ & $0.52\pm0.81$ & $0.78\pm1.14$ & $0.59\pm0.42$ & 3.37 & 31\\
\ion{O}{2} (3726 $\rm \AA$)  & 3726.0 & $4161.79\pm0.31$ & $0.72\pm0.09$ & $2.11\pm0.31$ & $4.25\pm0.68$ & 1.53 & 70\\
\ion{Ne}{3} (3869 $\rm \AA$) & 3869.0 & $4318.01\pm0.89$ & $0.57\pm1.33$ & $0.47\pm0.93$ & $0.56\pm0.47$ & 0.87 & 38\\
H$\beta$ (4861 $\rm \AA$) & 4861.0 & $5426.88\pm0.19$ & $0.64\pm0.09$ & $1.19\pm0.19$ & $1.92\pm0.55$ & 1.07 & 79 \\
\enddata
\tablecomments{Best-fit parameters for prominent emission lines in the optical spectrum of the host galaxy, derived using LIME. For each line, the table lists the rest-frame wavelength in vacuum, fitted line centre, amplitude, Gaussian width ($\sigma$), integrated line flux, and reduced $\chi^2$ of the fit. Uncertainties represent 1$\sigma$ errors from the fitting procedure. All fluxes are reported in units of $10^{-17}$ erg s$^{-1}$ cm$^{-2}$ \AA$^{-1}$. These fits are used to determine line diagnostics and redshift.}

\end{deluxetable}

\FloatBarrier

\begin{longtable*}{lllllll}
\caption{The extracted parameters from the equipartition analysis. All errors are $1\sigma$.\label{Table:equi_params}}\\
\hline
\hline
Geometry & Epoch & R ($10^{17}\:cm$) & E ($10^{50}\:erg$) & $\beta$ & B ($G$) & $n_o$ ($10^{-4}\:cm^{-3}$) \\
\hline
Spherical & 2023-05-04 & $4.68^{+0.33}_{-0.41}$ & $2.63^{+0.45}_{-0.36}$ & $0.288^{+0.015}_{-0.017}$ & $0.152^{+0.031}_{-0.026}$ & $475.62^{+75.36}_{-72.11}$ \\
 & 2023-07-07 & $4.93^{+0.30}_{-0.39}$ & $2.75^{+0.47}_{-0.37}$ & $0.272^{+0.012}_{-0.015}$ & $0.143^{+0.027}_{-0.024}$ & $424.35^{+54.76}_{-56.29}$ \\
 & 2023-08-27 & $5.49^{+0.26}_{-0.35}$ & $3.00^{+0.50}_{-0.40}$ & $0.275^{+0.010}_{-0.013}$ & $0.127^{+0.020}_{-0.019}$ & $336.77^{+26.94}_{-32.55}$ \\
 & 2023-11-29 & $6.12^{+0.23}_{-0.31}$ & $3.28^{+0.53}_{-0.43}$ & $0.266^{+0.008}_{-0.010}$ & $0.113^{+0.016}_{-0.015}$ & $265.98^{+11.99}_{-17.09}$ \\
 & 2024-05-05 & $6.79^{+0.23}_{-0.26}$ & $3.57^{+0.57}_{-0.49}$ & $0.246^{+0.006}_{-0.007}$ & $0.101^{+0.013}_{-0.012}$ & $212.50^{+6.44}_{-7.29}$ \\
 & 2024-06-24 & $7.45^{+0.26}_{-0.25}$ & $3.84^{+0.62}_{-0.56}$ & $0.252^{+0.007}_{-0.006}$ & $0.091^{+0.012}_{-0.010}$ & $173.96^{+5.71}_{-5.03}$ \\
 & 2024-09-28 & $7.99^{+0.28}_{-0.27}$ & $4.07^{+0.69}_{-0.61}$ & $0.245^{+0.006}_{-0.006}$ & $0.084^{+0.011}_{-0.010}$ & $149.20^{+5.94}_{-5.37}$ \\
 & 2024-12-26 & $8.51^{+0.31}_{-0.35}$ & $4.28^{+0.76}_{-0.65}$ & $0.241^{+0.007}_{-0.007}$ & $0.079^{+0.011}_{-0.010}$ & $130.18^{+6.76}_{-7.19}$ \\
 & 2025-03-20 & $8.98^{+0.38}_{-0.41}$ & $4.47^{+0.80}_{-0.69}$ & $0.237^{+0.008}_{-0.008}$ & $0.074^{+0.011}_{-0.010}$ & $115.83^{+8.32}_{-8.31}$ \\ \hline
Conical (5°) & 2023-05-04 & $84.58^{+5.73}_{-7.07}$ & $13.98^{+2.76}_{-2.16}$ & $0.880^{+0.008}_{-0.008}$ & $0.062^{+0.013}_{-0.011}$ & $36.52^{+5.20}_{-5.09}$ \\
 & 2023-07-07 & $89.16^{+5.29}_{-6.75}$ & $14.59^{+2.87}_{-2.24}$ & $0.871^{+0.007}_{-0.008}$ & $0.058^{+0.011}_{-0.010}$ & $32.59^{+3.70}_{-3.92}$ \\
 & 2023-08-27 & $99.21^{+4.52}_{-6.08}$ & $15.91^{+3.07}_{-2.42}$ & $0.873^{+0.005}_{-0.006}$ & $0.052^{+0.009}_{-0.008}$ & $25.86^{+1.68}_{-2.16}$ \\
 & 2023-11-29 & $110.66^{+3.95}_{-5.36}$ & $17.40^{+3.27}_{-2.63}$ & $0.868^{+0.004}_{-0.005}$ & $0.046^{+0.007}_{-0.006}$ & $20.43^{+0.63}_{-1.04}$ \\
 & 2024-05-05 & $122.76^{+3.87}_{-4.27}$ & $18.93^{+3.51}_{-2.98}$ & $0.855^{+0.004}_{-0.004}$ & $0.041^{+0.006}_{-0.005}$ & $16.32^{+0.26}_{-0.33}$ \\
 & 2024-06-24 & $134.63^{+4.28}_{-4.10}$ & $20.39^{+3.83}_{-3.33}$ & $0.859^{+0.004}_{-0.004}$ & $0.037^{+0.005}_{-0.005}$ & $13.36^{+0.25}_{-0.20}$ \\
 & 2024-09-28 & $144.50^{+4.63}_{-4.50}$ & $21.59^{+4.23}_{-3.64}$ & $0.855^{+0.004}_{-0.004}$ & $0.034^{+0.005}_{-0.004}$ & $11.46^{+0.29}_{-0.25}$ \\
 & 2024-12-26 & $153.85^{+5.25}_{-5.79}$ & $22.70^{+4.62}_{-3.87}$ & $0.852^{+0.004}_{-0.005}$ & $0.032^{+0.005}_{-0.004}$ & $10.00^{+0.37}_{-0.42}$ \\
 & 2025-03-20 & $162.37^{+6.47}_{-6.98}$ & $23.71^{+4.87}_{-4.11}$ & $0.849^{+0.005}_{-0.005}$ & $0.030^{+0.005}_{-0.004}$ & $8.89^{+0.51}_{-0.52}$ \\ \hline
Conical (15°) & 2023-05-04 & $29.92^{+2.04}_{-2.52}$ & $8.98^{+1.71}_{-1.34}$ & $0.722^{+0.014}_{-0.016}$ & $0.078^{+0.016}_{-0.014}$ & $72.21^{+10.58}_{-10.29}$ \\
 & 2023-07-07 & $31.54^{+1.89}_{-2.40}$ & $9.37^{+1.78}_{-1.39}$ & $0.705^{+0.013}_{-0.015}$ & $0.074^{+0.014}_{-0.013}$ & $64.43^{+7.56}_{-7.95}$ \\
 & 2023-08-27 & $35.10^{+1.62}_{-2.17}$ & $10.22^{+1.90}_{-1.50}$ & $0.708^{+0.010}_{-0.012}$ & $0.066^{+0.011}_{-0.010}$ & $51.13^{+3.52}_{-4.45}$ \\
 & 2023-11-29 & $39.15^{+1.42}_{-1.92}$ & $11.17^{+2.03}_{-1.64}$ & $0.699^{+0.008}_{-0.010}$ & $0.058^{+0.008}_{-0.008}$ & $40.38^{+1.39}_{-2.19}$ \\
 & 2024-05-05 & $43.43^{+1.39}_{-1.53}$ & $12.15^{+2.17}_{-1.85}$ & $0.676^{+0.007}_{-0.008}$ & $0.052^{+0.007}_{-0.006}$ & $32.26^{+0.64}_{-0.77}$ \\
 & 2024-06-24 & $47.63^{+1.54}_{-1.48}$ & $13.09^{+2.37}_{-2.08}$ & $0.683^{+0.007}_{-0.007}$ & $0.047^{+0.006}_{-0.006}$ & $26.41^{+0.59}_{-0.49}$ \\
 & 2024-09-28 & $51.12^{+1.66}_{-1.62}$ & $13.86^{+2.62}_{-2.27}$ & $0.675^{+0.007}_{-0.007}$ & $0.044^{+0.006}_{-0.005}$ & $22.65^{+0.66}_{-0.58}$ \\
 & 2024-12-26 & $54.43^{+1.88}_{-2.08}$ & $14.57^{+2.86}_{-2.42}$ & $0.670^{+0.008}_{-0.008}$ & $0.041^{+0.006}_{-0.005}$ & $19.76^{+0.81}_{-0.89}$ \\
 & 2025-03-20 & $57.45^{+2.32}_{-2.50}$ & $15.22^{+3.02}_{-2.56}$ & $0.665^{+0.009}_{-0.009}$ & $0.038^{+0.006}_{-0.005}$ & $17.59^{+1.07}_{-1.09}$ \\ \hline
Conical (30°) & 2023-05-04 & $18.22^{+1.23}_{-1.52}$ & $30.71^{+6.41}_{-5.00}$ & $0.612^{+0.017}_{-0.019}$ & $0.040^{+0.009}_{-0.007}$ & $10.90^{+1.48}_{-1.46}$ \\
 & 2023-07-07 & $19.21^{+1.14}_{-1.45}$ & $32.05^{+6.68}_{-5.19}$ & $0.593^{+0.015}_{-0.017}$ & $0.038^{+0.008}_{-0.007}$ & $9.72^{+1.04}_{-1.12}$ \\
 & 2023-08-27 & $21.37^{+0.97}_{-1.31}$ & $34.95^{+7.15}_{-5.59}$ & $0.596^{+0.011}_{-0.014}$ & $0.034^{+0.006}_{-0.005}$ & $7.71^{+0.45}_{-0.60}$ \\
 & 2023-11-29 & $23.84^{+0.85}_{-1.15}$ & $38.21^{+7.63}_{-6.10}$ & $0.586^{+0.009}_{-0.011}$ & $0.030^{+0.005}_{-0.004}$ & $6.09^{+0.15}_{-0.28}$ \\
 & 2024-05-05 & $26.44^{+0.83}_{-0.92}$ & $41.57^{+8.19}_{-6.89}$ & $0.559^{+0.008}_{-0.008}$ & $0.027^{+0.004}_{-0.003}$ & $4.87^{+0.05}_{-0.07}$ \\
 & 2024-06-24 & $29.00^{+0.92}_{-0.88}$ & $44.79^{+8.93}_{-7.69}$ & $0.567^{+0.008}_{-0.007}$ & $0.024^{+0.003}_{-0.003}$ & $3.99^{+0.05}_{-0.04}$ \\
 & 2024-09-28 & $31.13^{+0.99}_{-0.97}$ & $47.41^{+9.83}_{-8.38}$ & $0.559^{+0.008}_{-0.008}$ & $0.023^{+0.003}_{-0.003}$ & $3.42^{+0.07}_{-0.06}$ \\
 & 2024-12-26 & $33.14^{+1.12}_{-1.24}$ & $49.84^{+10.72}_{-8.91}$ & $0.553^{+0.009}_{-0.009}$ & $0.021^{+0.003}_{-0.003}$ & $2.98^{+0.09}_{-0.11}$ \\
 & 2025-03-20 & $34.98^{+1.39}_{-1.50}$ & $52.07^{+11.31}_{-9.44}$ & $0.547^{+0.010}_{-0.010}$ & $0.020^{+0.003}_{-0.003}$ & $2.65^{+0.13}_{-0.14}$
\\ \hline
\end{longtable*}

\clearpage

\bibliography{bibliography,luminosity_plot,Energy-velocity_plot}
\bibliographystyle{aasjournal}

\end{document}